\documentclass[aps,prb,twocolumn,keywords,nopacs,amssymb]{revtex4}

\usepackage[dvips]{graphicx}
\usepackage{dcolumn}
\usepackage{bm}
\usepackage{color}
\def\be{\begin{equation}}
\def\en{\end{equation}}
\def\bea{\begin{eqnarray}}
\def\ena{\end{eqnarray}}
\def\bec{\begin{equation}\begin{array}{rcl}}

\def\p{\partial}

\def\gs{\gtrsim}
\def\ls{\lesssim}
\def\kBT{k_{\rm B}T}
\def\ve{\varepsilon}
\newcommand{\av}[1]{\langle{#1}\rangle}

\newcommand{\bi}[1]{\mbox{\boldmath$#1$}}

\def\a1{\stackrel{\leftrightarrow}{1}}

\begin{document}
\title{Fluctuations of local electric field and dipole moments in water  
between metal walls 
 }  
\author{Kyohei Takae$^1$ and Akira Onuki$^2$}
\affiliation{
$^1$Institute of Industrial Science, University of Tokyo,
4-6-1 Komaba, Meguro-ku, Tokyo 153-8505, Japan\\
$^2$Department of Physics, Kyoto University, Kyoto 606-8502, Japan}
\date{\today}


\begin{abstract} 
We examine the thermal fluctuations of the local electric field 
${\bi E}_k^{\rm loc}$  and the dipole moment ${\bi \mu}_k$  in liquid water  
at $T=298~$K between metal walls  in   electric field applied   
in  the perpendicular direction. 
 We  use  analytic theory and molecular dynamics simulation. 
  In this situation,  there is a global electrostatic coupling 
between  the surface charges  on the  walls and   the polarization  
in the bulk.   Then, the  correlation function 
of the polarization density $ p_z({\bi r})$ along the  applied field 
contains a homogeneous part 
inversely proportional to the cell volume $V$.  
Accounting for the long-range dipolar interaction, 
we derive the   Kirkwood-Fr$\ddot{\rm{o}}$hlich   
formula for the polarization fluctuations 
when  the  specimen  volume  $v$ 
is much smaller than $V$.  However,  for not small $v/V$,  the  
homogeneous part comes into play   in dielectric relations. 
  We also calculate the  distribution of  
${\bi E}_k^{\rm loc}$  in applied field. 
As a unique feature of water, its magnitude $|{\bi E}_k^{\rm loc}|$ 
obeys  a Gaussian distribution with 
a large mean value $E_0 \cong 17~$V$/$nm, which arises 
 mainly from the surrounding hydrogen-bonded molecules. 
Since $|{\bi \mu}_k|E_0\sim 30 \kBT$, 
 ${\bi \mu}_k$ becomes mostly  parallel to ${\bi E}_k^{\rm loc}$. 
 As a result, the orientation distributions 
of these two vectors   nearly coincide, 
assuming the classical exponential form. 
In dynamics, the   component of ${\bi \mu}_k(t)$ 
parallel to  ${\bi E}_k^{\rm loc}(t)$  
changes on the timescale of the hydrogen bonds $\sim 5~$ps, 
while its smaller  perpendicular 
component   undergoes  librational  motions  
on timescales of 0.01 ps.  
\end{abstract}

\maketitle

\section{Introduction}

The dielectric 
properties of highly polar fluids are 
 complicated because of the long-range dipolar  
interaction. A great number of papers have been devoted on 
the dielectric constant $\ve$ in such fluids\cite{Fro}. 
 Lorentz and Debye presented classical formulas 
for  $\ve$ using  the concept of  
internal or local electric field acting on each dipole. 
However, their formulas were  inapplicable 
to  highly polar fluids, because they did   not 
properly account for the electrostatic interaction 
between a dipole considered and surrounding ones. 
Onsager\cite{Onsager} presented a mean-field theory, 
where  each  dipole is  
under influence of the reaction field produced by a 
surrounding continuum.  Kirkwood\cite{Kirk} further 
took  into account the short-range orientation 
correlation  introducing the 
so-called correlation factor $g_{\rm K}$. We mention subsequent 
statistical-mechanical theories\cite{Harris,Felder,Deu,Stell,Wer,Sm,Hansen,Netz}. 

In molecular dynamics simulations
\cite{Allen,Rah,Leeuw,Beren,Neu,Kusa,St,Yip,Raabe,Ab,Vega,Hansen,Netz}, 
$\ve$ has  been calculated  on the basis of  its 
 linear response  expressions   
 under   the three-dimensional (3D)  
periodic boundary condition. In such studies, 
     the Ewald method has been used   to 
efficiently sum  the electrostatic 
interactions \cite{Leeuw,Allen}.
Furthermore,  we mention simulations on  dipoles (and ions)  in  
 applied electric field 
  \cite{Hautman,Perram,Hender,Klapp,Takae,Madden1,Yeh,Voth,Takae1}. 
Some groups \cite{Hautman,Perram,Klapp,Takae,Takae1} introduced 
 image charges outside the cell 
to realize  constant electric potentials  
at $z=0$ and $H$ under   the periodic boundary 
condition in the $x$ and $y$ axes. In this paper, we use this method. 
As  simplified schemes of 
applying electric field,  Yeh and Berkowitz \cite{Yeh} 
assumed  empty slabs (instead of metal walls) 
in contact with water neglecting the image charges,   
while Petersen {\it et al.}\cite{Voth}   accounted  for
the primary image charges closest to the  walls. 
On the other hand, 
Willard {\it et al.}\cite{Madden1} 
used Siepmann  and Sprik's electrode model \cite{Sprik}, 
  where atomic particles form a
crystal in contact  with water  molecules 
and  their  charges vary continuously 
to realize the metallic boundary condition.

In this paper, we  examine the   pair  correlation function 
$G_{\alpha\beta}({\bi r}_1,{\bi r}_2)= 
\av{p_\alpha({\bi r}_1)p_\beta ({\bi r}_2)}$  
for the polarization density ${p}_\alpha({\bi r})$ 
($\alpha, \beta =x,y,z$).  For infinite systems,  
 Felderhof \cite{Felder} found that $G_{\alpha\beta}$
 consists of  short-range   and dipolar parts  and its 
  integral within a sphere  
  yields  the   Kirkwood-Fr$\ddot{\rm{o}}$hlich   
formula.  In our theory, there arises 
 a global electrostatic coupling between 
 the  fluctuations  of the  charge density 
on the metal  walls  and those of the  bulk 
polarization perpendicular to the walls (along the $z$ axis). 
As a result, the  $zz$ component  $G_{zz}$ contains 
a homogeneous part independent of ${\bi r}_1$ and ${\bi r}_2$ 
 due to  the presence of metal walls. This  homogeneous part is 
  inversely  proportional to the cell volume $V$ 
and largely alters  the fluctuations of 
 the total polarization $M_z=\int d{\bi r}p_z({\bi r})$ 
along the applied field. In our simulation, we 
also  demonstrate  the presence of the angle-dependent  
dipolar part decaying as $ |{\bi r}_1-{\bi r}_2|^{-3}$ 
in  $G_{\alpha\beta}({\bi r}_1,{\bi r}_2)$.

The local electric field acting on 
a  point dipole is the sum of the contributions  from the 
  other dipoles and 
the surface charges on the electrodes. 
For polar fluids,  the local field ${\bi E}_k^{\rm loc}$ on a molecule 
$k$ may be  defined as follows.  We  change 
  the molecular dipole ${\bi \mu}_k$   by $d{\bi \mu}_k$ 
and  express the resultant  change in the  electrostatic energy   as 
${\bi E}_k^{\rm loc}\cdot d{\bi \mu}_k$. 
In our recent paper on water\cite{Takae1}, we determined 
  ${\bi E}_k^{\rm loc}$  
in this  manner, which  is  a linear combination of the 
microscopic fields at the three molecular charge sites. 
In this paper,   we  investigate the microscopic basis 
of the dielectric properties of water  with this   expression.  

In the  classical  picture\cite{Fro}, 
 the local field is proportional to the 
applied field and is small. 
 However, in liquid water, 
 the microscopic  fields  on the molecular charges 
and the molecular local field  
are all  very large  being of order $15~$V$/$nm ($\sim e/\sigma^2$ 
with $\sigma=3.2~{\rm \AA}$)\cite{Sayka,Dellago,Ka1,Takae1}. 
As a unique feature of water, 
such strong electric  fields are mostly produced  
by the molecules hydrogen-bonded to each molecule $k$.  
In this paper, we  shall  see that 
the dipole  moment ${\bi \mu}_k$ is nearly 
parallel to the local field  ${\bi E}_k^{\rm loc}$ 
for each molecule even in applied field. 
In fact, if we  increase  the angle between ${\bi \mu}_k$ and 
${\bi E}_k^{\rm loc}$  
from 0 to $\pi/2$,  an energy of order 
$ \mu_0 |{\bi E}_k^{\rm loc}| \sim   30 \kBT$  is needed 
even at $T=298~$K,
where  $\mu_0=|{\bi \mu}_k|$.
Furthermore, our finding 
in applied field  is that  the distribution of  the  dipole  angle $\theta$ with respect to the applied  field  
 is of  the  exponential form  $\propto  \exp(\mu_0 E_{\rm p} \cos\theta/ \kBT)$ even in the nonlinear regime, where   
the effective field $E_{\rm p}$  yields the average polarization 
in accord with  the  Kirkwood-Fr$\ddot{\rm{o}}$hlich  theory in the linear 
regime.  Note that  this form was assumed in the classical 
theory\cite{Onsager,Kirk,Fro}, but 
 it needs to be  justified particularly   for polar fluids with strong 
local fields.

In dynamics, we shall see that each dipole ${\bi \mu}_k(t)$ consists of 
a  slow part on a timescale of 5 ps and a much smaller,  
fast part on a timescale of 0.01 ps. 
 The slow part is parallel to and moves 
with  ${\bi E}_k^{\rm loc}(t)$ , while 
the fast  part is perpendicular to ${\bi E}_k^{\rm loc}(t)$ 
and undergoes librational motions.    
This indicates that the dipole motions are  governed by 
the hydrogen bond dynamics in liquid water. 
In the previous simulations on liquid water 
\cite{Yip,Saito,Ohmine,Neu,Hynes,Laage,Kusa1,Gei},  
distinctly  separated 
 slow and fast molecular motions have been observed due to 
  the hydrogen bond network.

The organization of this paper is as follows. In Sec. II,
we will summarize the theoretical background 
and present  a theory on the polarization correlation function 
with electrodes.  
 In Sec. III, we will give numerical results 
on the thermal fluctuations of the local field and 
the dipole moments, the orientation time-correlation functions, 
 and the polarization correlations.  
In Appendix A, we will give an expression for the 
local electric field. 
In Appendix B, we will present a continuum theory of  
the polarization correlations to derive the 
homogeneous  part. In Appendix C, we will 
calculate  the polarization fluctuations 
in  a spherical specimen  under the influence of 
the reaction field. In Appendix D, we will 
calculate  the self-part 
of the polarization correlation function  
for water  
and the radius-dependent Kirkwood correlation 
factor.

\section{Theoretical  Background}

In this section, we explain our simulation  method 
   and present    analytic theory. Supplementary 
numerical results are also given.

\subsection{Electrostatics between metal walls}

As in our  previous work\cite{Takae1}, 
we use   the TIP4P$/$2005 model\cite{Vega} without ions, 
where  the  number of the water molecules is $N=2400$. 
 The temperature $T$ is fixed at $298$ K
in the $NVT$ ensemble with a 
Nos$\acute{\rm{e}}$-Hoover thermostat. 

For  each  molecule $k$,   
 its oxygen atom  and two protons are at 
 ${\bi r}_{k{\rm O}}$, ${\bi r}_{k{\rm H1}}$,
 and  ${\bi r}_{k{\rm H2}}$, respectively. 
Partial charges are  at  the protons and  another point 
${\bi r}_{k{\rm M}}$ with $q_{\rm H}=0.5564e$ and
 $q_{\rm M}= -2q_{\rm H}$, respectively. 
 The dipole moment of molecule $k$ is written as  
\be 
{\bi \mu}_k=q_{\rm H} ({\bi r}_{k{\rm H}1}+{\bi r}_{k{\rm H}2}- 2{\bi r}_{k{\rm M}}) 
= \mu_0 {\bi n}_k,
\en  
where ${\bi n}_k$ is the  unit vector along ${\bi \mu}_k$ 
with $\mu_0=2.305 {\rm D}$.  The charge position 
${\bi r}_{k {\rm M}}$ is  slightly shifted  from 
 ${\bi r}_{k {\rm O}}$ along ${\bi n}_{k}$.  There is no induced dipole moment 
in this model.  See Appendix A for more details. Hereafter, $k$ and $\ell$ 
represent molecules, while  $i=(k,\alpha)$ and $j=(\ell,\beta)$ stand for
molecular charges with $\alpha, \beta 
= {{\rm H}1}, {{\rm H}2},$  ${{\rm M}}$.

In our system, $N$ molecules  are in  
a $L\times L\times H$ cell 
with $L=41.5$${\rm \AA}$ and 
$H=44.7$$\rm \AA$, where smooth metal plates are at $z=0$ and $H$.   
 The periodic boundary condition 
is imposed along  the $x$ and $y$ axes. 
We apply electric field under the 
fixed-potential condition, 
where the electric potential 
$\Phi({\bi r})$ (defined away from the charge 
positions ${\bi r}\neq {\bi r}_j$)  satisfies   the metallic boundary condition,\be 
\Phi = 0\quad (z=0),\quad \Phi=-\Delta\Phi = -E_{\rm a} H  \quad (z=H),
\en 
where  $\Delta\Phi$ is the applied  potential difference 
  and  $E_{\rm a}=\Delta\Phi/H$   the applied  electric field.

To realize the condition (2), 
we introduce image  charges. 
For  each  charge $q_j$ at ${\bi r}_j=(x_j,y_j,z_j)$ in the cell, 
we consider   images with  the same  charge  $q_j$ at  
 $(x_j, y_j, z_j-2Hm_z)$  ($m_z= \pm 1, \cdots)$ 
and those with the opposite charge  $ -q_j $ at 
$(x_j, y_j, -z_j-2Hm_z)$ ($m_z=0, \pm 1, \cdots)$ outside the cell. 
Then, $\Phi({\bi r})$  is expressed as 
\be 
\Phi({\bi r})  =  \sum_{{\bi m}} {\sum_j}\bigg[
\frac{q_j}{|{\bi r}-{\bi r}_{j} + 
{\bi h}|} -   
\frac{q_j}{|{\bi r}-{\bar{\bi r}}_{j}  
 + {\bi h}|} \bigg]- E_{\rm a} z,
\en   
where ${\bi r}\neq {\bi r}_j$, 
${\bar{\bi r}}_j=(x_j,y_j, -z_j)$, and the  sum 
  is over ${\bi h}= (Lm_x,Lm_y,2Hm_z)$ 
with $m_x, m_y,$ and $m_z$ being integers. 
The sum over $m_x$ and $m_y$ arises from the lateral periodicity. 
Due to the sum over $m_z$, 
the first term in Eq.(3)  vanishes at  $z=0$ and $H$, leading to Eq.(2). 
The  electrostatic energy  
$U_{\rm m}$  is written as the sum \cite{Hautman,Perram,Klapp,Takae},  
\be
U_{\rm m} =\frac{1}{2}
 \sum_{\bi m} \bigg[ {\sum_{ij}}'
\frac{q_iq_j}{|{\bi r}_{ij} + 
{\bi h}|}  
-   {\sum_{ij}}
\frac{q_iq_j}{|{\bar{\bi r}}_{ij}  
 + {\bi h}|} \bigg] -E_{\rm a} M_z ,
\en  
where  
${{\bi r}}_{ij}={\bi r}_i- 
{{\bi r}}_j$,   and  ${\bar{\bi r}}_{ij}={\bi r}_i- 
{\bar{\bi r}}_j $.  In  $\sum_{ij}'$, we exclude the self term with 
$j=i$ for  ${\bi h}={\bi 0}$.  The $M_z$ is 
 the $z$ component of the 
 total polarization  ${\bi M}$, where 
\be 
{\bi M}=(M_x,M_y,M_z)=   \sum_k {\bi \mu}_{k}=\sum_i q_i {\bi r}_i.
\en 
We can apply  the 3D Ewald method 
due to the summation  over $\bi h$ fully accounting for the image charges 
\cite{Hautman,Perram,Klapp,Takae,Takae1}.

If  the  charge positions 
${\bi r}_i$ are shifted infinitesimally by  $d{\bi r}_i$ 
and  $E_{\rm a}$  by $dE_{\rm a}$,   $U_{\rm m}$ is changed  by     
\be
d U_{\rm m}  =  -\sum_i q_i {\bi E}_i \cdot d{\bi r}_i -M_z dE_{\rm a}, 
\en 
where ${\bi E}_i $ is  the microscopic  electric field acting on charge 
$i=(k,\alpha)$. Note that ${\bi E}_i$ consists of the contributions 
from the other molecules ($\ell\neq k$) for rigid water models.
 

The image charges are introduced as a mathematical 
convenience. The real charges are those in the cell and 
the surface charges  on the metal walls. 
We write the surface charge densities 
as  $\sigma_0(x,y)$ at $z=0$ and 
$\sigma_H(x,y)$ at $z=H$.
In our model,  there are no  charges  
in the thin regions $0<z\ls $1 $\rm\AA$  and $0<H-z\ls $1  $\rm\AA$  
due to the repulsive force  from the walls, so we can set $z=0$ and $H$ 
in Eq.(3) to find Eq.(2) 
(see a comment   below Eq.(12)). 
The surface charge densities  are  then  equal  to 
$ E_z(x, y, 0)/4\pi$ at $z=0$ and $-E_z(x, y,H)/4\pi$ 
at $z=H$,  where   $E_z =-\p \Phi/\p z$ is the electric field along the 
$z$ axis.  Their lateral averages 
${\bar \sigma}_\lambda=
\int dxdy~ \sigma_\lambda (x,y) /L^2$ ($\lambda=0, H$) 
are related to $M_z$ by \cite{Hautman,Takae1}  
\be 
{\bar{\sigma}}_0= -{\bar{\sigma}}_H 
=  E_{\rm a}/4\pi+  {M}_{z}/V .
\en  

The potential  from the  surface charges is divided into two parts as 
$  -4\pi {\bar \sigma}_0 z+ \phi_{\rm s}({\bi r}),
$ 
 where $\phi_{\rm s}$ arises from the deviations 
$\Delta\sigma_0=\sigma_0-{\bar \sigma}_0$ and    
$\Delta\sigma_H=\sigma_H-{\bar \sigma}_H$. 
The microscopic  field ${\bi E}_i$ in Eq.(6) is then divided  as  
\be 
{\bi E}_i={\bi E}_i^{\rm d}    + 
4\pi{\bar \sigma}_0 {\bi e}_z+ {\bi E}_{\rm s} ({\bi r}_i), 
\en 
where ${\bi e}_z$ is the unit vector along the $z$ axis. 
The ${\bi E}_i^{\rm d}$ arises from  the charges in the cell as   
\be  
{\bi E}_i^{\rm d} =  \sum_{ {\bi m}_\perp} 
 {\sum_{j}}'  { q_j}{\bi g}({{\bi r}_i- {\bi r}_{j} +  L{\bi m}_\perp}) ,  
\en 
where $ {\bi m}_\perp = (m_x, m_y,0)$ and 
  ${\bi g}({\bi r})= r^{-3} {\bi r}$. 
The amplitude  of ${\bi E}_i^{\rm d}$ is typically 
of order 15 V$/$nm  and is very large for water 
(see Sec.III). 
The third term ${\bi E}_{\rm s}({\bi r})=-\nabla\phi_{\rm s}({\bi r})$ 
decays  to zero  away from the walls\cite{Takae1}, but it grows 
as the charges approach the walls (the image effect).  
If we  neglect  ${\bi E}_{\rm s}$ 
retaining  $4\pi{\bar \sigma}_0 {\bi e}_z$ in Eq.(8), 
our method coincides with  that of 
Yeh and Berkowitz\cite{Yeh}.

The total potential $U$  consists of  three parts as 
\bea 
U&=&U_{\rm m}+  \frac{1}{2}\sum_{k\neq \ell}u_{\rm LJ}
(|{\bi  r}_{k{\rm O}}- {\bi r}_{\ell{\rm O}}|)\nonumber\\
&&\hspace{-5mm}+ \sum_{k} [u_{\rm w}(z_{k{\rm O}})+u_{\rm w}(H-z_{k{\rm O}})], 
\ena 
where $u_{\rm LJ}$ is the pair potential 
and $u_{\rm w}$ is the wall potential for the oxygen atoms 
expressed as 
\bea 
u_{\rm LJ}(r)&=& 4\epsilon[ (\sigma/r)^{12} - 
(\sigma/r)^{6}],\\
u_{\rm w}(z) &=& C_9 (\sigma/z)^9-C_3 (\sigma/z)^3.
\ena  
We set $\epsilon=93.2k_{\rm B}$,  $\sigma=3.1589{\rm \AA}$, 
$C_9=2\pi\epsilon/45$, and $C_3= 15C_9/2$. 
The density $n$ in liquid water is close to $\sigma^{-3}$ 
and $\sigma$ characterizes the molecular length. Also 
due to the repulsive part of $u_{\rm w}$, the molecules 
are depleted  next to the walls, which leads to Eq.(6). 

\subsection{Linear response and surface charge fluctuations}

The equilibrium distribution of our system 
is given by 
 ${\cal Z}^{-1} \exp(-{\cal H}/ \kBT)$, 
where the Hamiltonian ${\cal H}={\cal K}+U$ 
consists of  the  kinetic energy $\cal K$ 
and the potential $U$ in Eq.(10) and  
$\cal Z$ is the partition function. 
The  free energy is defined by $F(T, E_{\rm a}) = -\kBT \ln {\cal Z}$.  
From Eq.(4) $\cal H$ is expressed as   ${\cal H}={\cal H}_0 
-E_{\rm a}M_z$, where ${\cal H}_0$ is the 
Hamiltonian without applied field.
 Using this form, 
we may develop the linear response theory for small $E_{\rm a}$.  
In our theory, the symbol 
$\av{\cdots}$ denotes the average over this distribution. 
However, in our simulation results, 
  $\av{\cdots}$ denotes  
 the time   average over   6 ns.

As in spin systems, the average polarization 
$\av{M_z}$ and the polarization variance 
can be expressed as\cite{Onukibook} 
\be 
\av{M_z}=-\frac{\p F}{\p E_{\rm a}} ,\quad 
\av{(\delta M_z)^2}= {\kBT} 
\frac{\p \av{M_z}}{\p E_{\rm a}}, 
\en 
where $\delta M_z= M_z-\av{M_z}$ and the derivatives are performed at fixed 
$T$.  For any  physical quantity $\cal A$ (independent of $E_{\rm a}$), 
the thermal average $\av{{\cal A}}$ 
is a function of $T$ and $E_{\rm a}$ and its derivative with respect to $E_{\rm a}$ reads  
\be 
\frac{\p}{\p E_{\rm a}} \av{\cal A}=  \frac{1}{\kBT}  
 \av{{\cal A} \delta M_z }.
\en 
As   $E_{\rm a}\to 0$,  
 $\av{{\cal A}}$  is   expanded with respect to $E_{\rm a}$, 
leading  to the linear response expression,  
\be 
\av{{\cal A}}= \av{{\cal A}}_0+ \av{{\cal A} M_z}_0 E_{\rm a}/ \kBT +\cdots,
\en 
where  $\av{\cdots}_0$ is the equilibrium average with $E_{\rm a}=0$, 
so we have $ \av{M_z}_0=0$.
Using Eq.(7) we introduce  the effective dielectric constant 
$\ve_{\rm eff}$ of a film  by 
\be 
\ve_{\rm eff}=4\pi \av{\bar{\sigma}_0}/E_{\rm a} =1+4\pi \av{M_z}/VE_{\rm a},
\en 
where $V=L^2H$ is the film volume. As $E_{\rm a} \to 0$, we find 
\be
\lim_{E_{\rm a}\to 0}
\ve_{\rm eff}= 1+ {4\pi}\av{M_z^2}_0/V\kBT.
\en 
In our previous paper\cite{Takae1}, $\ve_{\rm eff}$ 
from Eq.(16) in the linear regime was $21.4$, 
which was close  to  that from  the right hand side of Eq.(17). 

\begin{figure}
\includegraphics[width=1\linewidth]{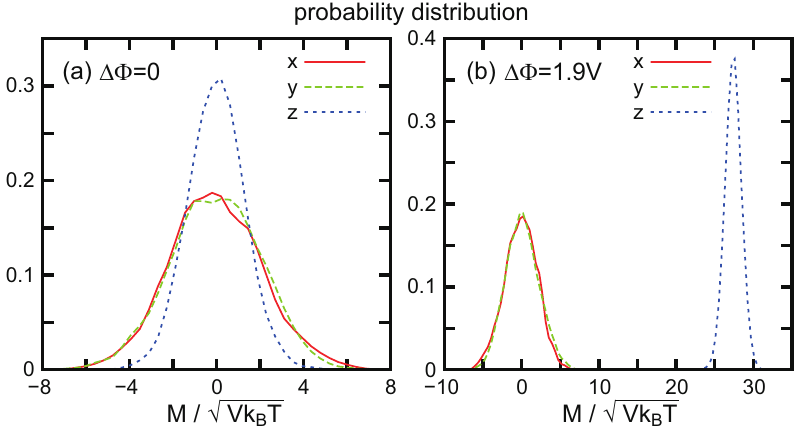}
\caption{ Gaussian distributions of 
$M_\alpha/(V\kBT)^{1/2}$ $(\alpha=x,y,z$) for 
$\Delta\Phi=0$ in (a) and 1.9 V in (b). 
Variance of $\delta M_z$ depends on $\Delta\Phi$ 
and is smaller than  those 
 of $M_x$ and $M_y$. 
 }
\end{figure}

In Fig.1, we show that  $M_z$, $M_x$, and $M_y$ 
obey Gaussian distributions. Here, 
$\av{(\delta M_z)^2}_0/V\kBT$ is $ 1.6$ and 1.2 
for $\Delta\Phi=0$  and 1.9 V, respectively, 
while    $\av{M_x^2}_0/V\kBT=\av{M_y^2}_0/V\kBT \cong 4.4$   
both for $\Delta\Phi=0$ and 1.9 V. 
Here, if $\Delta\Phi=1.9~$V, the applied 
field is $E_{\rm a}=0.42~$V$/$nm. 
Thus, the fluctuations 
of $M_z$ are considerably suppressed  than those 
of $M_x$ and $M_y$. 

The surface charge densities consist 
 of  the deviation  and the mean value. 
That is, we write $\sigma_0={\bar\sigma}_0+\Delta\sigma_0(x,y)$  at $z=0$. 
The  mean   ${\bar\sigma}_0$ gives rise to 
a homogeneous electric field $4\pi {\bar \sigma}_0 {\bi e}_z$ 
in the cell  as in Eq.(8). From Eq.(7), 
  ${\bar\sigma}_0$  is a fluctuating variable with 
variance  \cite{Limmer}  
\be 
\av{(\delta{\bar\sigma}_0)^2}= \frac{1}{V^2} 
\av{(\delta M_z)^2} =\frac{\kBT}{V^2} 
\frac{\p \av{M_z}}{\p E_{\rm a}},
\en 
where $\delta{\bar\sigma}_0= {\bar\sigma}_0-\av{{\bar\sigma}_0}$ 
is the deviation from the thermal average $\av{{\bar\sigma}_0}$. 
 Thus,  $\av{(\delta{\bar\sigma}_0)^2}L^2 \to 
\kBT(\ve_{\rm eff}-1)/4\pi H$ as  $E_{\rm a}\to 0$, which  
 decreases  as  $H^{-1}$  for  large $H$. 
See Appendix B for 
more discussions. In contrast,  the deviation $\Delta\sigma_0(x,y)$ 
is mostly  determined by   the molecules   close to the wall\cite{Takae1} 
and   the electric field ${\bi E}_{\rm s}({\bi r})$ 
in Eq.(8) is noticeable  only close to the wall. 
Recently, Limmer {\it et al.} discussed the surface charge 
fluctuations\cite{Limmer}.

\subsection{Equilibrium states under applied field}
In equilibrium under applied  field,  Stern layers  with a microscopic 
thickness  $d$ appear near the walls, where 
a  potential drop occurs. However, dielectric response  depends on 
another longer  length $\ell_{\rm w}$ in highly polar fluids.  
In this paper, the cell width $H$ is much longer than 
$d$ but  can be shorter than $\ell_{\rm w}$.

\subsubsection{Stern layers and bulk polarization}

We  introduce a microscopic polarization density ${\bi p}({\bi r})$ 
in terms of the charge positions ${\bi r}_j$ for polar molecules. 
It  is related to the  microscopic charge density  
$\rho({\bi r})$ by \cite{Takae1}
\be 
-\nabla\cdot{\bi p} = \rho = \sum_j q_j \delta ({\bi r}-{\bi r}_j). 
\en 
See Eq.(A4) in Appendix A for the expression of ${\bi p}$. 
The total polarization (5) is given by the integral 
${\bi M}= \int d{\bi r}{\bi p}({\bi r})$ in the cell.  
In our geometry, the average polarization $P(z)\equiv  \av{p_z({\bi r})}$ 
along the $z$ axis depends only on $z$.  Using  this $P(z)$  and the average 
surface charge density $\av{{\bar{\sigma}}_0}$ we   define  
the  average Poisson  potential $\Psi(z)$   by  
\be
\Psi(z)=- 4\pi \av{{\bar{\sigma}}_0} z + 4\pi \int_0^z dz' P(z'),
\en   
Here,  $\Psi(0)=0$, while  $\Psi(H)= -HE_{\rm a}$ 
follows from Eq.(7). 
The other groups  \cite{Hautman,Yeh,Madden1} determined  
$\Psi(z)$  from  $d^2\Psi/dz^2= -4\pi\av{\rho}(z)$ 
without introducing $\bi p$.  
For general geometries without ions,  the 
 average potential $\Psi({\bi r})$  is obtained from 
  $\nabla^2\Psi=4\pi \nabla\cdot\av{{\bi p}({\bi r})}$ with 
boundary conditions.

In our case, 
 Stern layers are formed 
even for pure water with thickness about $5~\rm\AA$, where 
the molecules are under strong influence of the walls. 
In experiments, the Stern  layers  have been  observed 
on ionizable dielectric walls 
in aqueous solutions\cite{Beh,Bazant}, where 
the layers can be influenced by ion adsorption 
 and/or  ionization on  the surface. 

Outside the  layers,  a homogeneously polarized state 
is realized \cite{Hautman,Yeh,Madden1,Takae,Takae1,Hender,Voth} 
  with thickness $H-2d$, where    $P(z)\cong P_{\rm b}$ and 
$-d\Psi(z)/dz\cong E_{\rm b}$ with 
\be 
4\pi\av{{\bar\sigma}_0}= E_{\rm b} + 4\pi P_{\rm b}=\ve E_{\rm b},
\en  
where $\ve$ is  the bulk dielectric constant. 
In  Fig.2(a), we show  $P_{\rm b}/n\mu_0$ vs. $\Delta \Phi$.  
Here, $n$ is the  average density  in the bulk 
($n=1.071/\sigma^{3}$  for $\Delta\Phi=0$ 
and $n=1.058/\sigma^{3}$ 
 for $\Delta\Phi=1.9$ V).  We can see that 
 $P_{\rm b}$  increases  linearly for  $\Delta\Phi\ls 4$ V 
but it tends to saturate for larger $\Delta\Phi$.  
In our previous paper\cite{Takae1}, we 
 determined   $\ve$ directly 
from $\ve = 1+ 4\pi P_{\rm b}/E_{\rm b}$ to 
obtain  $\ve\sim 60$  for  $\Delta\Phi \ls 4~$V, 
but this method is not accurate for very small $P_{\rm b}$ and 
$E_{\rm b}$.

\subsubsection{Surface electric length }

In terms of a surface  electric length  $\ell_{\rm w}$ \cite{Takae1}, 
the  sum of the potential drops in the top and bottom Stern layers is 
 given by $\Delta\Phi\ell_{\rm w}/(\ell_{\rm w} +H)$, so 
the potential decreases by  
 $\Delta\Phi H/(\ell_{\rm w} +H)$ in the bulk.  
 As a result, the bulk quantities 
$\ve$ and  $E_{\rm b}$ are related to  the 
film quantities $\ve_{\rm eff}$ and  $E_{\rm a}$ by   
\be
\ve/\ve_{\rm eff}=E_{\rm a}/E_{\rm b} = 1+\ell_{\rm w}/H. 
\en 
In  Fig.2(b), we plot   $\ell_{\rm w}$ vs. $\Delta\Phi$, 
which is calculated from Eq.(26) below (so its accuracy is not 
 good for very small $\Delta\Phi$).  Here, 
 $\ell_{\rm w}\sim 9$ nm for $\Delta\Phi\ls 4~$V. 
In the  previous simulations\cite{Hautman,Yeh,Madden1}, 
$\ell_{\rm w}\sim 2H$. In our case, as $E_{\rm a}\to 0$, 
we have  $\ell_{\rm w}/H \cong 1.86$, 
$\ve/\ve_{\rm eff}=E_{\rm a}/E_{\rm b} \cong 2.86$, and $\ve\sim 60$. 

From Eqs.(17) and (22), the variance of  $M_z$  at $E_{\rm a}=0$ 
is written as 
\be 
\av{M_z^2}_0 =  \frac{V\kBT}{1+\ell_{\rm w}/H} \bigg(\chi- \frac{\ell_{\rm w}}{4\pi H}
\bigg) \cong  \frac{V\kBT\chi}{1+\ell_{\rm w}/H},
\en 
where we can omit 
 $\ell_{\rm w}/  4\pi H(\ll  \chi)$ for water. 
Hereafter, 
\be 
\chi=(\ve-1)/4\pi
\en 
  is the bulk susceptibility. 
  On the other hand, 
$M_x$ and $M_y$  are decoupled from ${\bar\sigma}_0$ (see Eqs.(33) and 
(37)). Thus,  
\be 
\av{M_x^2}_0 =\av{M_y^2}_0= V\kBT\chi.  
\en

\begin{figure}
\includegraphics[width=1\linewidth]{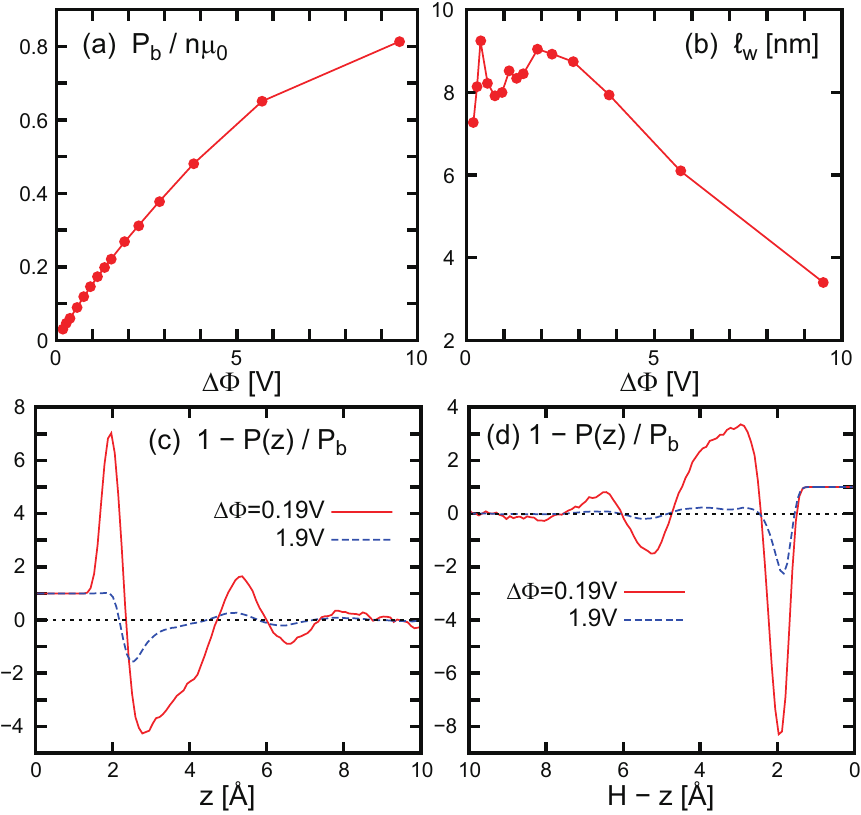}
\caption{ (a) Normalized bulk polarization density  $P_{\rm b}/n\mu_0$ 
vs. $\Delta\Phi$, where $n\mu_0$ is the maximum 
polarization. (b) Surface electric length 
  $\ell_{\rm w}$ vs. $\Delta\Phi$.  Displayed also are 
profiles of $1-P(z)/P_{\rm b}$ at $\Delta\Phi=0.19 $ 
and $1.9$ V  near  the bottom  in (c) 
 and  the top in (d). Its integral is $\ell_{\rm w}/(\ve-1)$ from 
Eq.(26).  Due to depletion, 
$P_{\rm b}\cong 0$  for $z<1.5~{\rm \AA}$ and  $H-z<1.5~{\rm \AA}$. 
For larger separation, variations  for 
$\Delta\Phi=0.19~$V are  amplified 
than those for $\Delta\Phi=1.9~$V because of smaller $P_{\rm b}$.
 }
\end{figure}

We  express  $\ell_{\rm w}$ in terms of $P(z)$. 
From Eqs.(20) and (22) we obtain 
 $(E_{\rm a}-E_{\rm b})H= 4\pi\int_0^Hdz [P_{\rm b}-P(z)]$, so   
\be 
\ell_{\rm w}
= (\ve-1) \int_0^H dz [1-P(z)/P_{\rm b}], 
\en 
where  the integrand  $1-P(z)/P_{\rm b}$  
is nonvanishing only around the Stern layers. 
Hence,  $\ell_{\rm w}$  is  independent of $H$ for 
$H\gg d$.  It is worth noting that the integration  
 of $1-P(z)/P_{\rm b}$ 
is analogous to the calculation  of a Gibbs dividing 
  surface between two coexisting phases\cite{Netz}.

In Fig.2(c) and (d), we display  $1-P(z)/P_{\rm b}$  for 
$\Delta\Phi=0.19$ and 1.9 V near the bottom  and top. 
 Its  integral is  of order $1~{\rm \AA}$ mainly due to the  depletion 
 at the walls, but $\ell_{\rm w}$ is 
about 9 nm  due to large  $\ve-1$. 
 For $\Delta\Phi=0.19~$V, 
 we can see a  maximum at $z\sim 2~{\rm \AA}$ in (c) 
and a minimum  at $H-z\sim 2~{\rm \AA}$ 
in (d), which originate from  preference of the H-down  
configurations due to the image effect\cite{Takae1}. However,  
these extrema disappear for   $\Delta\Phi=1.9~$V due to the influence of 
the increased surface charge density given by   
$\av{{\bar{\sigma}}_0}/e= 0.44/$nm$^2$. Oscillatory behavior 
follows for larger separations. The curves of $\Delta\Phi=0.19~$V 
and 1.9 V are very different, but their integrals 
are nearly the same as in Fig.2(b).   

\subsubsection{Surface  capacitance}

The  potential change  
in the Stern layers can be calculated from the Poisson potential 
$\Psi(z)$ in Eq.(20)  as
\bea 
(\Delta\Phi)_{\rm S} &=&  \av{{\bar\sigma}_0}/C_+ +\Phi_{00}^{\rm w} \quad 
(z\cong  0),\nonumber\\
&=& -\av{{\bar\sigma}_H}/C_- -\Phi_{00}^{\rm w} \quad (z\cong H), 
\ena   
where $\av{{\bar\sigma}_H}=- \av{{\bar\sigma}_0}$  
and $C_+$  and $C_-$ are  the surface capacitances 
\cite{Hautman,Yeh,Madden1,Takae1,Hender}.   
Due to  the molecular anisotropy of  water,  
there  appears an intrinsic   potential change at zero applied field 
given by  
$ 
\Phi_{00}^{\rm w}=-4\pi \int_0^d dz P(z),
$  
which   sensitively depends on the 
wall potential \cite{Hautman,Madden1,Takae1}.  
In fact,  $\Phi_{00}^{\rm w}$  was $0.2~$V  in Willard {\it et al.}'s 
simulation\cite{Madden1}, but it was $ -0.09~$V for weaker adsorption 
in our simulation \cite{Takae1}.

  The sum of these potential drops  
gives the  total boundary drop   
 in  the form  ${\bar \sigma}_0/C$ with  
\be 
 C=(C_+^{-1}+C_-^{-1})^{-1}, 
\en 
where $\Phi_{00}^{\rm w}$ does not appear. 
In terms of this $C$, the length $\ell_{\rm w}$ is expressed as 
\be 
\ell_{\rm w}=\ve/4\pi C  .
\en 
It is worth noting that   the surface capacitance 
of  a dielectric film on a metal wall 
is given by  $C_d= \ve_d/4\pi \ell_d$, 
where $\ve_d$ is its dielectric constant and  $\ell_d$ is its   thickness.

\subsection{Polarization  correlations}

\subsubsection{Dipolar and homogeneous  correlations}

We consider the polarization correlation function,
\be 
G_{\alpha\beta}({\bi r}_1, {\bi r}_2)=
 \av{\delta p_\alpha({\bi r}_1) 
\delta p_\beta({\bi r}_2)} ,
\en 
where   $\alpha,\beta=x,y,z$ and 
$\delta p_\alpha=  p_\alpha-\av{ p_\alpha}$. 
In this subsection,  $E_{\rm a}$ is  small 
 in the linear response regime, so $\av{\cdots}\cong 
\av{\cdots}_0$ in Eq.(30) and 
 $\av{(\delta M_\alpha)^2}\cong \av{M_\alpha^2}_0$.

 Felderhof\cite{Felder} presented a continuum 
theory of the polarization fluctuations for infinite,  uniform  
  systems without  electrodes.   For  isotropic  polar systems, he 
found short-ranged and dipolar correlations as        
\be 
{G_{\alpha\beta}({\bi r}_1, {\bi r}_2)}/ {\kBT}= \chi  \delta_{\alpha\beta} 
\delta ({\bi r})  + ({\chi}^2/{\ve}) \nabla_\alpha \nabla_\beta r^{-1},
\en 
where ${\bi r}={\bi r}_2-{\bi r}_1$  and $
\nabla_\alpha=\p/\p x_\alpha$. Here, $\nabla_\alpha\nabla_\beta r^{-1}= 
3x_\alpha x_\beta/{r^5}- {\delta_{\alpha\beta}}/r^3$, 
which  diverges as $r\to 0$ and is meaningful 
only for $r\gs \sigma$.   
 The delta function in Eq.(31) 
should also be treated as a normalized 
shape function with a width longer than $\sigma$. 
We may conveniently remove this  divergence of the dipolar term  
by replacing    $r^{-1} $ by its long-range part, 
 written as $\psi_\ell(r)$.  In the Ewald method, 
use has been   made of the form \cite{Allen,Leeuw}, 
\be 
\psi_\ell(r)={\rm erf} (\gamma r)/r, 
\en    
where $\gamma\sim\sigma^{-1}$ and 
 ${\rm erf}(u) = (2/\sqrt{\pi})\int_0^u du e^{-u^2}$ is the error 
function.  Here, $\psi_\ell\cong r^{-1}$ for $r\gg \gamma^{-1}$ 
and $\psi_\ell =(2\gamma/\sqrt{\pi})(1- \gamma^2r^2/3\cdots )$  for 
 $r\ll \gamma^{-1}$. 
 Notice that  the inner product 
 $\sum_\alpha G_{\alpha\alpha}
= \av{\delta {\bi p}({\bi r}_1)\cdot  
\delta {\bi p}({\bi r}_2)}$ has no  long-range dipolar 
correlation.
The dipolar correlation itself 
follows in the  continuum theory  
(from  Eq.(B1) in Appendix B),  
as in the case of dipolar ferromagnets\cite{Ah}.

In our theory, Eq.(7) indicates that 
the fluctuations of the mean surface charge density 
  ${\bar\sigma}_0$ produce polarization 
fluctuations homogeneous in the bulk along the $z$ axis (see Appendix B). 
We   propose  the following form,   
\bea 
{G_{\alpha\beta}({\bi r}_1, {\bi r}_2)}/ {\kBT}&=& \chi  \delta_{\alpha\beta} 
\delta ({\bi r})  + ({\chi^2}/{\ve}) 
 \nabla_\alpha \nabla_\beta \psi_\ell(r) \nonumber\\
&&+ \delta_{\alpha z} \delta_{\beta z}  {h_{\rm o}}/{V},
\ena 
where ${\bi r}_1$ and ${\bi r}_2={\bi r}_1+{\bi r}$ 
are in the bulk  and $h_{\rm o}$ is a constant to be determined in Eq.(39). 
We have  replaced 
  $r^{-1}$ by $\psi_\ell$ and 
neglected deformation of the dipolar correlation 
 due to  the metal walls.  
If we pick up two molecules $k$ and $\ell$ in the bulk, 
their correlation at large separation 
($\gg  \sigma$) tends to a constant as  
\be 
\av{{\bi \mu}_k\cdot{\bi \mu}_\ell}=\mu_0^2\av{{\bi n}_k\cdot{\bi n}_\ell}
\cong \kBT h_{\rm o} /n^2 V,
\en 
where the dipolar correlation vanishes. 

Setting ${\bi r}_1= {\bi r}_0+{\bi r}'$ and ${\bi r}_2= 
{\bi r}_0+ {\bi r}$ in Eq.(33), we   integrate $G_{\alpha\alpha}=
G_{\alpha\alpha}({\bi r}_0+{\bi r}',{\bi r}_0+ {\bi r})$ 
($\alpha=x,y,z$)    with respect to ${\bi r}=(x,y,z)$ 
within a spheroid  expressed by 
\be 
 (x^2+y^2)/R_\perp^2 + z^2/R_\parallel^2 \le 1,
\en 
where  the radii $R_\perp$ and  $R_\parallel$ 
much exceed $\gamma^{-1}$   and  ${\bi r}'$ is also 
within the spheroid (away 
 from its surface). The  two points 
${\bi r}_0+{\bi r}'$ and ${\bi r}_0+ {\bi r}$ are in the bulk region. 
 These integrations 
  can  safely be performed  because the dipolar terms 
$\nabla_\alpha\nabla_\beta \psi_\ell(|{\bi r}-{\bi r}'|)$ are  finite 
at  $|{\bi r}-{\bi r}'|=0$.  The resultant integrals  are  independent of
 ${\bi r}'$   as\cite{surface} 
\bea 
&&\hspace{-5mm}\int_{\rm spheroid}\hspace{-2mm}
 d{\bi r} {G_{zz}}/{\kBT} = \chi- \frac{4\pi}{\ve}\chi^2  
{N_z}  +h_{\rm o}\frac{ v}{V}  , \\
&&\hspace{-5mm}\int_{\rm spheroid}\hspace{-2mm}
 d{\bi r} G_{xx}/{\kBT}  = \chi- \frac{2\pi}{\ve}\chi^2  
({1-N_z})   ,
 \ena   
where  $v= 4\pi R_\parallel R_\perp^2/3$ is    the spheroid volume.  
 The integral of ${G_{yy}}$ is equal to  that of ${G_{xx}}  $.  The $N_z$ is 
the depolarization factor along the $z$ axis 
determined by \cite{Landau}
\be 
\eta \equiv  R_\parallel/R_\perp. 
\en 
The $N_z$  tends to   1 for $\eta \ll 1$, 
$1/3$ for $\eta = 1$, and 0 for  $\eta \gg 1$. 
 The last term  in Eq.(36) arises from the 
homogeneous term in Eq.(33), so it is proportional to  $v$.

We also  integrate $G_{zz}({\bi r}_1, {\bi r}_2)$ 
in Eq.(33) with respect to ${\bi r}_1$ and $ {\bi r}_2$ 
 in the whole  bulk region, where we can use Eq.(36)  
in the pancake limit    $N_z=1$ 
under the lateral periodic boundary condition \cite{comment2}.  
 In Appendix B, 
we shall also see that the polarization sum in the 
Stern layers is much smaller than that 
 in the bulk region. Thus,    we have   
$
\av{(\delta M_z)^2}/\kBT=V(\chi/\ve+ h_{\rm o}).
$  
 From Eq.(23)  we now determine   $h_{\rm o}$ as 
\be
h_{\rm o} =  \chi/(1+ \ell_{\rm w}/H)- \chi/\ve=(\ve_{\rm eff}-1)\chi/\ve,
\en    
which considerably depends on $H$  for $H \ls \ell_{\rm w}$ but  tends 
to $ 4\pi\chi^2/\ve$ for $H\gg \ell_{\rm w}$. Another derivation of 
Eq.(39)   will  be presented   in Appendix B. 
For $N_z=1$ in Eq.(36), the first short-range term and 
the second dipolar term almost cancel for $\ve \gg 1$, but 
the third homogeneous term increases with increasing $v$ 
leading to Eq.(23) for $v=V$ (see Fig.14).
We  also obtain  Eq.(25) from Eq.(37) with $N_z=1$, where 
 the dipolar part does not contribute.

\subsubsection{Kirkwood-Fr$\ddot{o}$hlich   formula}


To reproduce Kirkwood-Fr$\ddot{\rm{o}}$hlich   formula, 
we  define   the polarization 
integrals $\delta M_\alpha^{\rm s} =
\int_{r<R} d{\bi r}\delta p_\alpha({\bi r}_0+{\bi r})$ 
within a sphere with radius $R$. 
Then, 
 we find 
\be 
  \av{ (\delta { M}_\alpha^{\rm s})^2}/v \kBT 
 ={\chi(2\ve+1)}/{3\ve}  +\delta_{\alpha z} h_{\rm o}{ v}/{V}  . 
\en 
Without the second term, 
  the above relation  
is well-known in the literature\cite{Fro,Kirk,Onsager}. 
For a sphere in infinite systems, 
the  factor $(2\ve+1)/3\ve$ in the first term was shown to 
arise from  the   reaction field produced 
by  the molecules in the  exterior\cite{Onsager}. 
In Appendix C, we will present another simple 
derivation  of   Eq.(40) in the limit $v/V\to 0$. 
 In Fig.3,  $\delta M_\alpha^{\rm s}$ obey Gaussian distributions. 
For $R=0.95~$nm, 
$ \av{ (M_\alpha^{\rm s})^2}\cong 2.9v\kBT$ for the three 
directions consistently 
 with Eq.(40). 
However, the boundary effect appears 
for $R=1.58$ nm, where $ \av{ (M_z^{\rm s})^2}\cong 2.3v \kBT$ 
and  $ \av{ (M_x^{\rm s})^2}= \av{ (M_y^{\rm s})^2}\cong 3.2v \kBT$.


\begin{figure}
\includegraphics[width=1\linewidth]{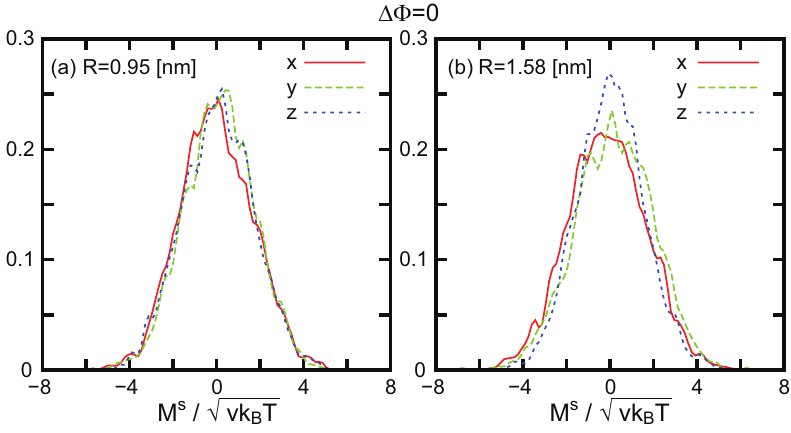}
\caption{ Gaussian distributions of 
$M_\alpha^{\rm s}/(v\kBT)^{1/2}$ $(\alpha=x,y,z$) in a spherical cavity, whose  
radius $R$ is (a) $0.95~$nm  and (b) 1.58 nm  at $\Delta\Phi=0$. 
In (a), there is almost no difference in the three directions 
with  variance $\av{(M _\alpha^{\rm s})^2}\cong 2.9v \kBT$. 
In (b), it is decreased to 
$ 2.3v \kBT$ for $\alpha=z$ 
and is increased to 
 and  $ 3.2v \kBT$ for $\alpha=x,y$ 
due to the boundary effect.   
 }
\end{figure}

The left hand side of Eq.(40) 
has  been  written as $n\mu_0^2 g_{\rm K}/\kBT$ for infinite systems 
($v/V \to 0$) 
in terms of the   correlation factor $g_{\rm K}$, which leads to 
  the  Kirkwood-Fr$\ddot{\rm{o}}$hlich   formula 
for nonpolarizable molecules\cite{Kirk,Fro},
\be 
  \chi  =({n\mu_0^2 g_{\rm K}}/{\kBT}) {\ve}/({2\ve +1}).  
\en 
The Onsager formula\cite{Onsager}  follows for $g_{\rm K}=1$. 
The factor $g_{\rm K}$ is  defined in terms of 
the dipole correlation 
between molecules $k$ and $\ell$ in the bulk  as\cite{Kirk}  
 \be
g_{\rm K}=1+ {\sum_{ \ell }}' 
\av{{\bi n}_k\cdot{\bi n}_\ell},
\en 
where  the first term 1 arises 
from the self term ($\ell=k$) and  the  sum  in the second term is over  
$\ell (\neq k)$  at fixed $k$ with 
 molecule $\ell$ being in a sphere   with volume $v\ll V$. 
Here,  the dipolar correlation vanishes 
and  the nearest neighbor molecules  give a dominant  
contribution in  the  sum\cite{Kirk,Fro}. This  assures that 
the formula (41) holds in the thermodynamic limit.  
This has been confirmed  
by many authors\cite{Rah,Neu,Kusa,Yip,Beren}.

In our case, we   obtain $g_{\rm K}=2.14$ from Eq.(42) 
at $\Delta\Phi=0$ (see Fig.15(b) in Appendix D),  
which yields $\chi=4.68$ and $\epsilon=59.8$ 
from Eq.(41).  Now, in addition to this derivation, 
we have determined $\ve$  from the dielectric relation (21)\cite{Takae1}  
and from  the data of $\av{M_z^2}_0$ and $\ell_{\rm w}$ 
(see the sentence below Eq.(22)), 
all  yielding $\ve\cong 60$ 
in the linear regime.  A close value of 
 $\ve$ was obtained  by Abascal and Vega\cite{Vega} 
for the TIP4P$/$2005 model  under 
the 3D periodic boundary condition.

It is worth noting that 
  Harris and Alder\cite{Harris}  added 
 the term  $n\alpha_{\rm m} (\ve+2)/3$ 
in the right hand side of Eq.(41) 
for polarizable fluids, where   $\alpha_{\rm m}$ 
is  the molecular polarizability. Their expression  
reduces to the Clausius-Mossotti formula $3\chi/(\ve+2)=n\alpha_{\rm m}$ 
as  $\mu_0 \to 0$. It was used 
to  analyze 
a wide range of data on $\ve$ for  water\cite{Sengers}. 

\section{Numerical results}

\begin{figure}
\includegraphics[width=0.96\linewidth]{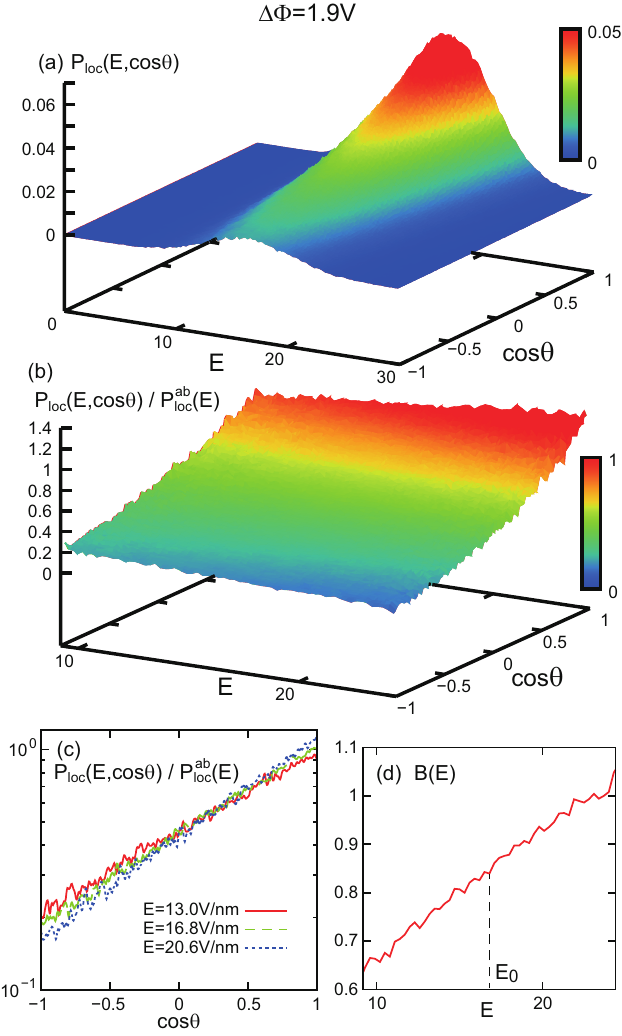}
\caption{   (a) Joint distribution $P_{\rm loc}(E,\cos\theta)$ 
of the local electric field for   $E= |{\bi E}_k^{\rm loc}|$ 
and  $\cos\theta= {E}_{zk}^{\rm loc}/|{\bi E}_k^{\rm loc}|$ in Eq.(44), where  $\Delta\Phi=1.9~$V. 
 (b) Conditional probability distribution 
$P_{\rm loc}(E,\cos\theta)/P_{\rm loc}^{\rm ab}(E)$ 
and  (c) its logarithm for $E=13.0, 16.8$, and 20.6 V$/$nm, where 
$P_{\rm loc}^{\rm ab}(E)$ is the distribution for 
$E= |{\bi E}_k^{\rm loc}|$  in Eq.(46). 
This  conditional distribution is of  
  the exponential form $\propto 
\exp(B\cos\theta)$.  (d)  $B$  depends on $E$ linearly.  
 }
\end{figure}

\begin{figure}
\includegraphics[width=1\linewidth]{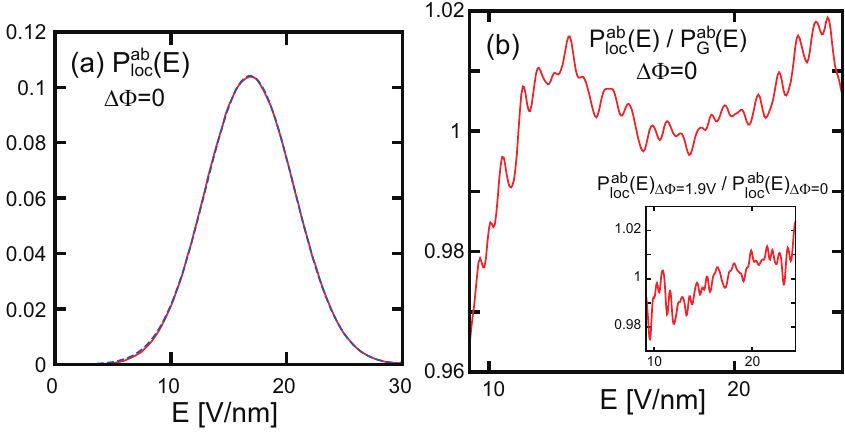}
\caption{ (a)  Distribution $P_{\rm loc}^{\rm ab}(E)$ 
for  $E= |{\bi E}_k^{\rm loc}|$  in Eq.(46) at $\Delta\Phi=0$ (red) 
and Gaussian distribution  $P_{\rm G}^{\rm ab}(E)$ in Eq.(50) (blue). 
The two curves nearly coincide. 
 (b)    Ratio of this distribution 
to the Gaussian one ($=P_{\rm loc}^{\rm ab}(E)
/P_{\rm G}^{\rm ab}(E))$ at $\Delta\Phi=0$. 
Shown also is  the ratio 
of the distribution at $\Delta\Phi=1.9~$V 
to that at $\Delta\Phi=0$ ($=
P_{\rm loc}^{\rm ab}(E)_{\Delta\Phi=1.9{\rm V}}
/P_{\rm loc}^{\rm ab}(E)_{\Delta\Phi=0}$) (inset). They are 
 very close to 1 in the displayed range $9~$V$/$nm $<E<25$ V$/$nm. 
}
\end{figure}

In this section, 
we further present  results of our molecular dynamics  simulation 
on the fluctuations of the local fields  and the dipoles.

\subsection{Fluctuations of local field} 

Electric  field  fluctuations 
have been numerically examined in water   and  electrolytes 
(for $\Delta\Phi=0$)\cite{Sayka,Dellago,Ka1,Takae1}, because 
they play  crucial roles  in dissociation reactions 
and vibrational spectroscopic response.
Here, we  examine the thermal fluctuations of the 
local electric field ${\bi E}_k^{\rm loc}$,  which is  defined  by Eq.(A2) 
in Appendix A.  

\subsubsection{Joint distribution}

For each molecule $k$, the direction of ${\bi E}_k^{\rm loc}$ is  written as  
\be 
{\bi e}_k=  |{\bi E}_k^{\rm loc}|^{-1}
{\bi E}_k^{\rm loc}.
\en 
In terms of its $z$ component ${ e}_{zk}$, 
the angle of the local field with respect to the $z$ axis 
is given by $\cos^{-1}({ e}_{zk})$. We consider the joint distribution for 
 $|{\bi E}_k^{\rm loc}|$ and ${ e}_{zk}$:  
\be 
P_{\rm loc}(E, \cos\theta)= \av{
\delta(|{\bi E}_k^{\rm loc}| -{ E})
\delta({ e}_{zk} -\cos\theta) }_{\rm b},
\en 
which is related to the 3D local-field distribution as 
\bea 
P_{\rm loc}^{\rm 3D}({\bi E}) 
&=&  \av{\delta({\bi E}_k^{\rm loc}-{\bi E})}_{\rm b} \nonumber\\
&=& ( 2\pi E^2)^{-1}  P_{\rm loc}(E, \cos\theta).
\ena    
Hereafter, $\av{\cdots}_{\rm b}=  \av{
\sum_{k\in {\rm bulk}}N_{\rm b}^{-1}\cdots }$ denotes 
 the average   over  the  molecules 
in the  region $0.3H<z_{k{\rm G}}<0.7H$ 
and over a time interval of 6 ns, 
where $z_{k{\rm G}}$ is the $z$ component of 
the center of mass  of molecule $k$  
and $N_{\rm b}$ is the number of the molecules in this  region. 
The distribution of $ |{\bi E}_k^{\rm loc}|$ and 
that of  $e_{zk} $  are  separately written as   
\bea
&&\hspace{-9mm}
P_{\rm loc}^{\rm ab}(E) 
=\int_{-1}^1 ds P_{\rm loc}(E, s)
=\av{\delta(|{\bi E}_k^{\rm loc}|-E)}_{\rm b},  
\\
&&\hspace{-9mm} 
P_{\rm loc}^{\rm or}(s) 
=\int_{0}^\infty \hspace{-2mm}
 dE  P_{\rm loc}(E, s)
 = \av{\delta( e_{zk} -s)}_{\rm b} , 
\ena
where we write $s=\cos\theta$.

In  Fig.4,  we display  (a) $P_{\rm loc}(E, \cos\theta)$ 
and (b) the conditional probability  distribution 
$P_{\rm loc}(E, \cos\theta)/P_{\rm loc}^{\rm ab}(E)$ 
in the $E$-$\cos\theta$ plane at $\Delta\Phi=1.9~$V.  
The former is peaked at an intermediate $E_0$ 
 independent of $\cos\theta$, while the latter 
 depends on $E$ very weakly in the displayed range of $E$.  
In (c), the latter is nearly of the exponential form,  so   
\be 
{P_{\rm loc}(E,\cos\theta)}\cong{ P_{\rm loc}^{\rm ab}(E)}
 \frac{\exp[B(E) \cos\theta]}{Z(B(E))}, 
\en
where $Z(x)= 2\sinh x/ x$ is the normalization factor.  
The coefficient  $B=B(E)$ weakly  depends on $E$  as 
\be 
 B(E) \cong B(E_0)  +0.46 (E/E_0-1) ,
\en 
where $E_0$ gives a maximum of ${ P_{\rm loc}^{\rm ab}(E)}$ 
(see Eq.(50)).

\subsubsection{Large local-field amplitude due to hydrogen bonding}

\begin{figure}
\includegraphics[width=1\linewidth]{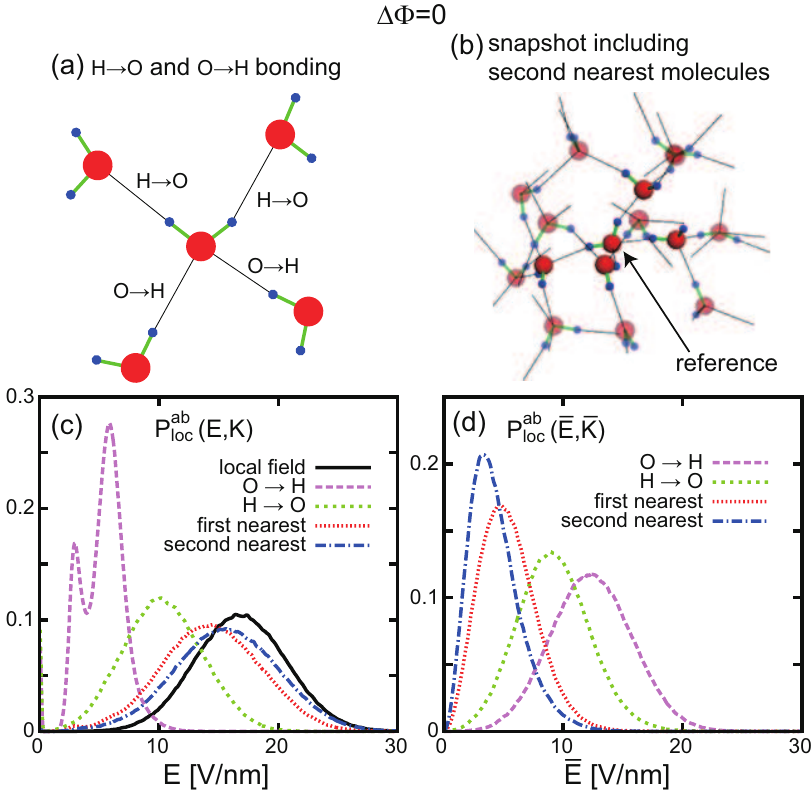}
\caption{ (a) Typical configuration of 4 hydrogen-bonded 
molecules (first-nearest ones) 
around  a reference molecule at the center. 
Hydrogen bonds (black bars) are from 
oxygen (hydrogen) of the reference one  to 
hydrogen (oxygen) of the other ones, written as O$\to$H (H$\to$O). 
(b) Snapshot of  hydrogen-bonded molecules 
composed of 4 first-nearest ones and 11 second-nearest ones.
(c) Distribution $P_{\rm loc}^{\rm ab}(E,K)$ in Eq.(53)  for  local-field 
contributions   from  group $K$: 
(i) O$\to$H, (ii) H$\to$O, (iii) 
first(O$\to$H+H$\to$O), and (iv) first+second ones. They 
 are compared with $P_{\rm loc}^{\rm ab}(E)$. 
(d) Distribution $P_{\rm loc}^{\rm ab}({\bar E},{\bar K})$ in Eq.(54) 
for  local-field  contributions from group $\bar K$, 
where the contributions from 
group $K$ are excluded.  
 }
\end{figure} 

In  Fig.5,  $P_{\rm loc}^{\rm ab}(E)$  
 is almost independent of  $\Delta\Phi$ in the linear response regime
and is very close to   a  shifted Gaussian distribution expressed as  
\be
P_{\rm G}^{\rm ab}(E)
 \equiv   \frac{1}{\sqrt{2\pi s_0}}
\exp\bigg[- \frac{(E-E_0)^2}{2s_0}\bigg] ,
\en
whose  mean and  standard deviation are given by 
\bea 
&&\hspace{-5mm} E_0=16.8 {\rm V}/{\rm nm}=1.2e/\sigma^2=31\kBT/\mu_0,\\
&&\sqrt{s_0}=3.8 {\rm V}/{\rm nm} =  0.23 E_0.
\ena  
Sellner {\it et al}.\cite{Ka1}  calculated 
 $P_{\rm loc}^{\rm ab}(E)$  at O and H sites 
for $\Delta\Phi=0$ to find similar behavior.

In water, it is natural that 
 a large fraction of  the local field is produced by 
the hydrogen-bonded molecules around  each  molecule. 
 Reischl {\it et al.}\cite{Dellago} analyzed this aspect 
at  the centers of the molecules' OH bonds. 
Note that there are a number 
of definitions of the hydrogen bonds\cite{Luzar,Kumar,Rao,Zie}. 
As in our previous  paper\cite{Takae1}, 
we treat two molecules to be hydrogen-bonded
if one of the intermolecular OH distances is shorter than 2.4 $\rm \AA$
and the angle between the OO vector and one of their
intramolecular OH bonds is smaller than $\pi/6$. A similar
definition was used by Zielkiewicz\cite{Zie}.
  The average number  
of the intermolecular hydrogen bonds is then 3.6  per molecule in the bulk. 
For   this  HO-distance 
definition,  the hydrogen bonds for each molecule k   
extend  either  from its  protons (H$\to$O) 
or from its  oxygen atom (O$\to$H). 
This classification is convenient   for the present problem.

In Fig.6, we display a  typical configuration of 4  hydrogen-bonded 
molecules (first nearest ones) in (a) 
and  a snapshot of  15 ones  (first and second nearest ones) in (b) 
around a reference molecule at the center.
The second  nearest ones are those   hydrogen-bonded 
to the first nearest ones\cite{comment3}. 
These surrounding ones give rise 
 ${\bi E}_k^{\rm loc}$ which is  nearly parallel to   
${\bi \mu}_k$  (see Fig.9 below). 
In Fig.6(c),  we write the distributions, 
\be   
P_{\rm loc}^{\rm ab}(E,K)= \av{
\delta(|{\bi E}_k^{\rm loc}(K)| -{ E})}_{\rm b}.
\en 
For each reference $k$,  ${\bi E}_k^{\rm loc}(K)$ are 
the  local-field contributions  from the other 
molecules in group $K$, 
where  $K$ represents  
  (i) O$\to$H, (ii) H$\to$O, (iii) 
first(O$\to$H+H$\to$O), and (iv) first+second ones. 
To $P_{\rm loc}^{\rm ab}(E)$, 
the distribution from the first nearest ones 
is considerably  close 
and that from the first+second ones is 
very close. Also  the local-field contributions from the  
nearest H$\to$O bonded  molecules are most important. This 
 is because   ${\bi E}_k^{\rm loc}$ 
is rather close to  $\frac{1}{2}({\bi E}_{k{\rm H1}}+{\bi E}_{k{\rm H2}})$ 
(see Eqs.(A2) and (A3) in Appendix A). 
 Furthermore, in Fig.6(d), we display  the local-field contributions 
from  group $\bar K$ which consists of the 
molecules not belonging to   group $K$.  
Their distributions are  written as   
\be   
{P}_{\rm loc}^{\rm ab}({\bar E},{\bar K})= \av{
\delta(|{\bi E}_k^{\rm loc}-{\bi E}_k^{\rm loc}(K)| -{\bar E})}_{\rm b}.
\en 
Naturally,  the distributions 
excluding the first nearest neighbors 
and the first+second ones 
have smaller mean values and variances.

\subsubsection{Orientation distribution of local field }

In Fig.7(a), the angle distribution  
 $P_{\rm loc}^{\rm or}(\cos\theta)$   is 
excellently fitted to  the exponential  form, 
\be 
P_{\rm loc}^{\rm or}(\cos\theta)\cong  
\exp(B_{\rm loc}\cos\theta)/Z(B_{\rm loc}).  
\en 
The coefficient $B_{\rm loc}$ is related to $B(E)$ in Eq.(48) by 
\be 
B_{\rm loc} \cong  B(E_0).
\en 
The above form also follows 
 from integration of Eq.(48) with respect to $E$ 
with the aid of  Eqs.(49) and (50), where we use the sharpness 
of $P_{\rm loc}^{\rm ab}(E)$ or the inequality $s_0/E_0^2 \ll 1$.   
See Fig.8(a) for 
$B_{\rm loc}$ vs. $\Delta\Phi$, where the linear growth 
($\propto \Delta\Phi$) occurs for $\Delta\Phi \ls 4~$V as in  
 Fig.2(a).  From Eqs.(48), (50), and (55) 
the bulk average of the $z$ component of the local field is written as 
\be 
E_{\rm loc}\equiv \av{E_{zk}^{\rm loc}}_{\rm b}=E_0 \av{e_{zk}}_{\rm b}=  
E_0 {\cal L}(B_{\rm loc}), 
\en 
where ${\cal L}(x)= \coth (x)-1/x$ is the Langevin function. 
In the linear regime $B_{\rm loc}\ll 1$, 
we have $E_{\rm loc}\cong E_0 B_{\rm loc}/3$.

We may also consider the distribution 
of the local field in one  direction,    
which was calculated previously 
\cite{Sayka,Dellago,Ka1,Takae1}. 
It is defined by 
$P_{\rm loc}^{\rm 1D}(E_z)=\av{\delta (E_{zk}^{\rm loc}-E_z)}_{\rm b}$ 
along the $z$ axis.   From Eqs.(45), (48), (50), and (55), 
we obtain 
\be
P_{\rm loc}^{\rm 1D}(E_z)\cong 
\frac{\exp(B_{\rm loc}E_z/E_0)}{Z(B_{\rm loc})} 
 \int_{|E_z|}^\infty \hspace{-2mm} 
dE' \frac{ P_{\rm loc}^{\rm ab}(E')}{2E'}.
\en 
For $\Delta\Phi=0$, this distribution 
 exhibits a plateau in the  range $|E_z|< E_0$  from Eq.(50)\cite{Ka1,Takae1}.
For  $\Delta\Phi=1.9~$V, its profile was calculated in  our previous paper
\cite{Takae1}.

\begin{figure}
\includegraphics[width=1\linewidth]{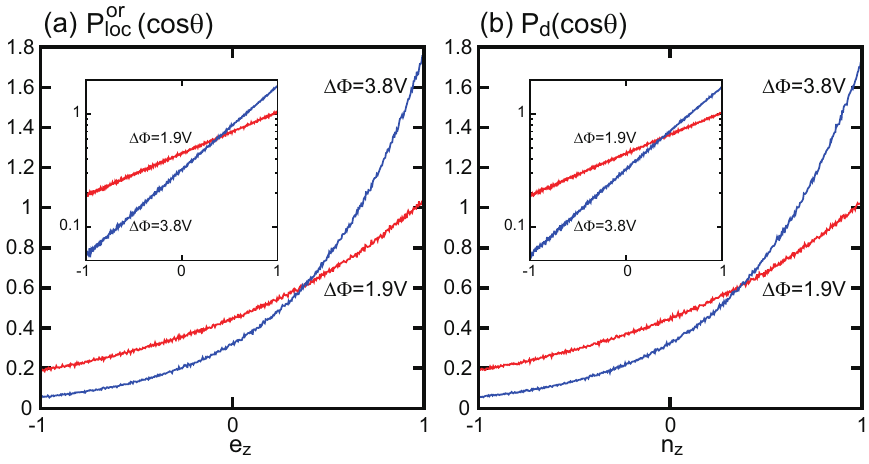}
\caption{  Distribution $P_{\rm loc}^{\rm or}(\cos\theta)$ 
for the local field orientation   $\cos\theta= {e}_{zk}$ (left) 
and that $P_{\rm d}(\cos\theta)$ 
for the dipole orientation  $\cos\theta= { n}_{zk}$ (right) for 
$\Delta\Phi=1.9$ and 3.8 V. They nearly coincide 
and their logarithms  are linear in $\cos\theta$  (inset), leading to 
the exponential forms in Eqs.(55) and (59) 
with $B_{\rm loc}=0.85 (1.71)$ and  $B_{\rm d}=0.84 (1.68)$
for $\Delta\Phi=1.9~(3.8)~$V. 
 }
\end{figure}

\begin{figure}
\includegraphics[width=1\linewidth]{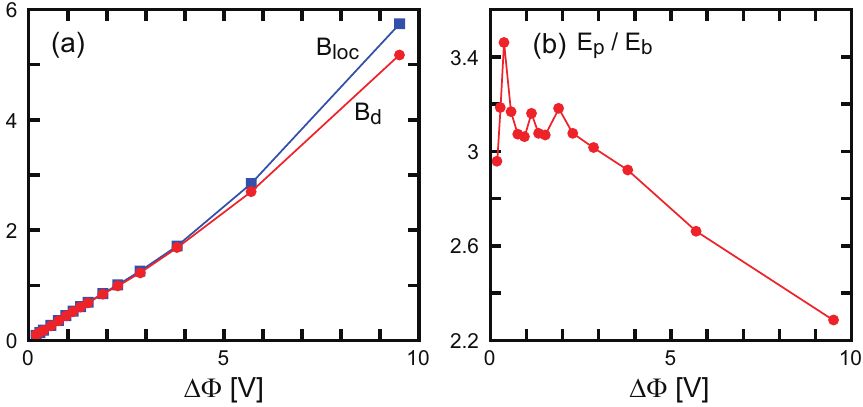}
\caption{ (a) Coefficients $B_{\rm loc}$ and $B_{\rm d}$ in the 
exponential orientation distributions in Eqs.(55) and (59) 
vs. $\Delta\Phi$. (b) 
$E_{\rm p}/E_{\rm b}$ vs. $\Delta\Phi$, where $E_{\rm p}$ is the effective 
field proportional to $B_{\rm d}$ 
in  Eq.(61) and $E_{\rm b}$ is the bulk  field in Eq.(21).  
 }
\end{figure}

\subsection{Dipole  fluctuations}

\subsubsection{Classical exponential form}

For  the polarization direction ${\bi n}_k= \mu_0^{-1} {\bi\mu }_{k}$ 
we consider  the distribution of its  $z$ component 
$n_{zk}=\cos\theta_k$.  In Fig.7(b), it is again 
nearly of the exponential form,  
\bea
 P_{\rm d} (\cos\theta) &=&  \av{\delta( n_{zk}-\cos\theta)}_{\rm b} \nonumber\\
& \cong & \exp( B_{\rm d} \cos\theta )/Z(B_{\rm d} ).   
\ena 
This  exponential form was  assumed 
in  the classical  theories\cite{Onsager,Kirk,Fro}. 
Recently, it was   numerically obtained for water  
by  He {\it et al.}\cite{Koga}.
Furthermore,  in Fig.7, $P_{\rm d}(\cos\theta)$ 
 is very close to $P_{\rm loc}^{\rm or}(\cos\theta)$ 
for the local field orientation in Eq.(55). 
In Fig.8(a),  the difference 
between $ B_{\rm loc}$ and $ B_{\rm d}$ 
is indeed very small. In terms of  $B_{\rm d}$,  
  the average bulk  polarization $P_{\rm b}$ in Eq.(21)  is expressed as 
\be 
P_{\rm b}= n\mu_0\av{n_{zk}}= n\mu_0 {\cal L}(B_{\rm d}). 
\en 
Here, we introduce the effective electric field $E_{\rm p}$ by \cite{Onsager} 
\be 
E_{\rm p}= \kBT B_{\rm d} /\mu_0=  (\kBT/\mu_0) {\cal L}^{-1} (P_{\rm b}/n\mu_0),
\en 
where ${\cal L}^{-1}(x)$ is the inverse function  of ${\cal L}(x)$. 
In the linear response regime,  we obtain \cite{Fro} 
\be
E_{\rm p}= 3\kBT P_{\rm b}  /n\mu_0^2= [3\ve g_{\rm K} /(2\ve+1)]E_{\rm b} .
\en
In Fig.8(b), we plot $E_{\rm p}/E_{\rm b}$ using the data in Fig.2(a), 
which  is about 3 in the linear regime but decreases considerably 
in the nonlinear regime.


Since $ B_{\rm loc}\cong  B_{\rm d}$, Eqs.(57) and (60) give 
\be 
 E_{\rm loc}/P_{\rm b} \cong E_0 /n\mu_0\cong 7.0. 
\en 
Here, we introduce  
the Lorentz factor $\gamma_{\rm loc}$   by 
\be  
E_{\rm loc}= E_{\rm b}+ 4\pi\gamma_{\rm loc}P_{\rm b}. 
\en 
From Eq.(63), $\gamma_{\rm loc} $ 
can be expressed in terms of $E_0$   as 
\be 
\gamma_{\rm loc}\cong ( E_0/\mu_0 n-1/\chi)/4\pi,  
\en 
which gives $\gamma_{\rm loc}
\cong 0.56$  even in the nonlinear regime. 
In  our previous paper\cite{Takae1}, 
 we directly calculating $E_{\rm loc}=\av{E_{zk}^{\rm loc}}_{\rm b}$ 
to obtain  $\gamma_{\rm loc}
\cong 0.58$ for any $\Delta\Phi$.  
Here,  if we set $P_{\rm b}=n(\alpha_{\rm m}+ \mu_0^2/3\kBT)E_{\rm p}$ 
with $E_{\rm p}=  E_{\rm loc}$ and 
 $\gamma_{\rm loc}=1/3$, we obtain 
 the Debye formula\cite{Fro,Sengers} 
 $3\chi/(\ve+2) =n(\alpha_{\rm m}+\mu_0^2/3\kBT)$, 
where $\alpha_{\rm m}$ is the molecular polarizability. 
 For liquid water, however, we find 
 $E_{\rm loc}/E_{\rm p}\cong  \mu_0E_0/3\kBT \cong 10$ from 
 Eqs.(62) and (63)  in the linear regime.

\begin{figure}
\includegraphics[width=1\linewidth]{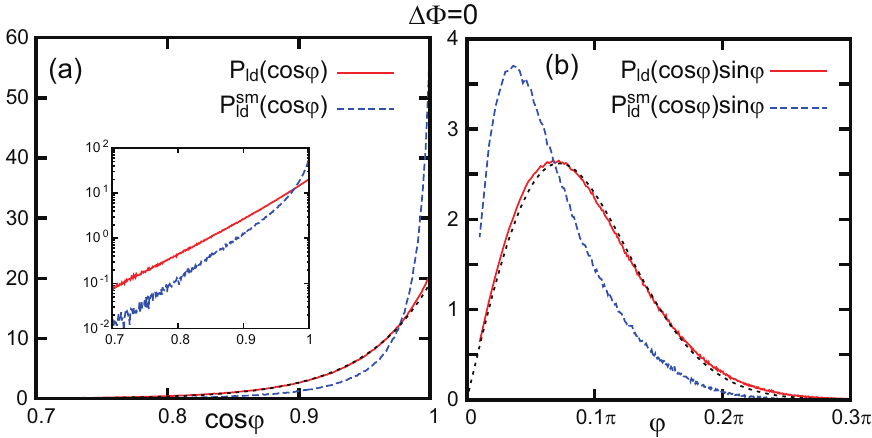}
\caption{ (a) Distribution $P_{\rm ld}(\cos\varphi)$ 
for $\cos\varphi= {\bi n}_k\cdot{\bi e}_k$  in Eq.(66) 
and $P_{\rm ld}^{\rm sm}(\cos\varphi)$ 
for $\cos\varphi= {\bar {\bi n}}_k\cdot{\bi e}_k$ in Eq.(68) 
at $\Delta\Phi=0$,  
where  ${\bar{\bi n}}_k$ is the time-averaged polarization direction 
in Eq.(67). Their logarithms are also shown (inset). 
Then,  $P_{\rm ld}(\cos\varphi)
\propto \exp(19.6\cos\varphi)$, but   
$P_{\rm ld}^{\rm sm}(\cos\varphi)$ 
does not exhibit the exponential form for small $\varphi$. 
(b)  $ P_{\rm ld}(\cos\varphi)\sin\varphi$ and 
$ P_{\rm ld}^{\rm sm}(\cos\varphi)\sin\varphi$ vs. $\varphi$ at $\Delta\Phi=0$. 
 }
\end{figure}

\subsubsection{  Nearly parallel ${\bi n}_k$ and ${\bi e}_k$  }

The  result  $P_{\rm d} (\cos\theta) \cong  P_{\rm loc}^{\rm or}(\cos\theta)$ 
in applied field in Fig.7 indicates that 
the dipole direction    ${\bi n}_k$  and  the local field direction 
${\bi e}_k$ should be  nearly parallel both without and with applied field, 
despite large thermal fluctuations at $T=298~$K.   As discussed 
in Sec.I, this is because of   large $\mu_0 E_0/\kBT\sim 30$ in Eq.(51). 
We thus consider the angle 
$\varphi_k=\cos^{-1}({\bi n}_k\cdot{\bi e}_k)$ 
between the two directions. 
 In Fig.9(a), its  distribution at $\Delta\Phi=0$ behaves as  
\bea 
P_{\rm ld}(\cos\varphi)& =& 
\av{\delta (\cos\varphi-  {\bi n}_k\cdot{\bi e}_k)}_{\rm b} \nonumber\\
&\cong &  \exp( B_{\rm ld} \cos\varphi )/Z(B_{\rm ld} ), 
\ena 
with  $B_{\rm ld}=19.6$. We obtain   $\av{\cos\varphi_k}_{\rm b}=0.95$ 
and $\av{(\cos\varphi_k)^2}_{\rm b}=0.90$, so    $\sqrt{\av{\varphi_k^2}_{\rm b}} 
\sim 0.3 \sim 0.1 \pi$.   In  Fig.9(b), we also plot 
$P_{\rm ld}(\cos\varphi) \sin\varphi$ 
vs. $\varphi$, which is approximated as   $B_{\rm ld}\varphi \exp(- B_{\rm ld}
\varphi^2/2)$ for small $\varphi$.   This distribution is 
insensitive to $\Delta\Phi$ (not shown here). 

Moreover,  as will be  shown below, 
${\bi n}_k(t)$ contains a fast librational  part 
on timescales of order 0.01 ps, while  ${\bi e}_k(t)$ 
changes on timescales of order 5 ps. 
Thus, we  define the temporally smoothed  polarization direction, 
\be 
{\bar{\bi n}}_k (t) ={\cal N}_k^{-1}  \int_{-\Delta t/2}^{\Delta t/2}
 \hspace{-2mm} d\tau~ {\bi n}_k(t+\tau), 
\en 
where we set  $\Delta t=0.03~$ps and ${\cal N}_k$
is the normalization factor   ensuring 
$|{\bar{\bi n}}_k|=1$ with $|\av{{\cal N}_k^2}_{\rm b}|^{1/2}  
\cong 0.9\Delta t$.  The angle 
${\bar \varphi}_k=\cos^{-1}({\bar{\bi n}}_k\cdot{\bi e}_k)$ 
for  ${\bar{\bi n}}_k $ should be  smaller than $\varphi_k$.  In Fig.9(a), 
we  plot its angle  distribution,  
\be 
P^{\rm sm}_{\rm ld}(\cos\varphi) =\av{\delta (\cos\varphi-  
{\bar{\bi n}}_k\cdot{\bi e}_k)}_{\rm b}, 
\en 
which is surely sharper than $P_{\rm ld}(\cos\varphi)$ in Eq.(66) 
with   $\av{\cos{\bar \varphi}_k}_{\rm b}=0.97$ 
and  $\av{(\cos{\bar \varphi}_k)^2}_{\rm b}=0.94$.   
 In  Fig.9(b), the curve of 
$P^{\rm sm}_{\rm ld}(\cos\varphi) \sin\varphi$ 
vs. $\varphi$ is more sharply peaked 
for  small $\varphi$  than $P_{\rm ld}(\cos\varphi) \sin\varphi$.


\begin{figure}
\includegraphics[width=1\linewidth]{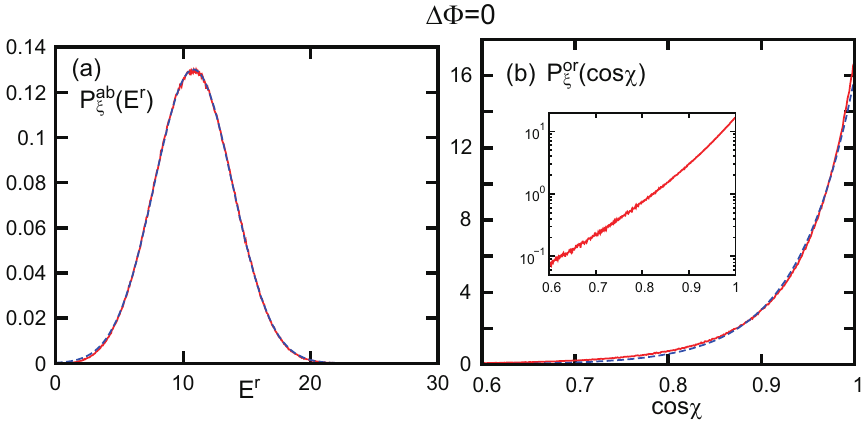}
\caption{ Distributions for   vector ${\bi \xi}_k$ 
between the two protons in Eq.(69)  and its conjugate force 
${\bi F}_k^{\rm r}$ in Eq.(70).    (a)  $P_\xi^{\rm ab}(E^{\rm r})$ in Eq.(71)   
for the amplitude  $q_{\rm H}^{-1}|{\bi F}_k^{\rm r}|$ 
 at $\Delta\Phi=0$, which is  Gaussian 
form with mean    $10.8~$V$/$nm and standard deviation $3.1~$V$/$nm. 
(b) $P_\xi^{\rm or}(\cos\chi)$ in Eq.(72) 
for the angle between ${\bi \xi}$ 
and ${\bi F}_k^{\rm r}$  at $\Delta\Phi=0$, where 
 these two directions are nearly  parallel. 
Shown also is its logarithm (inset),  indicating 
 the  form $\propto \exp(B_\xi \cos\theta)$ with $B_\xi\cong 16.6$.
} 
\end{figure}

\subsection{Relative vector between two protons}

For each  molecule $k$, we may also consider 
the relative vector between the two protons,
\be 
{\bi\xi}_k ={\bi r}_{k{\rm H1}}-{\bi r}_{k{\rm H2}},
\en  
whose length is fixed at $ |{\bi \xi}_k| =a_{\rm HH}=1.514~{\rm \AA}$. 
From Eq.(A1) in Appendix A, its  conjugate force  is given by 
\be 
{\bi F}_k^{\rm r} = \frac{1}{2} 
{q_{\rm H}}({\bi E}_{k{\rm H1}}-{\bi E}_{k{\rm H2}}). 
\en 

As in the dipole case, we first consider 
 the distribution for the field amplitude 
 $|{\bi F}_k^{\rm r}|/q_{\rm H}= |{\bi E}_{k{\rm H1}}-{\bi E}_{k{\rm H2}}|/2$:
\be 
P_\xi^{\rm ab}( E^{\rm r})= \av{\delta (E^{\rm r}- |{\bi F}_k^{\rm r}|/q_{\rm H})}_{\rm b} .
\en  
In Fig.10(a), $P_\xi^{\rm ab}( E^{\rm r})$ is of    the  
Gaussian form in Eq.(50). Its  mean value  and standard deviation  
are  $10.8~$V$/$nm and 
$3.1~$V$/$nm, respectively. This form  indicates 
strong   asymmetry between 
 the forces acting on the two protons due to the hydrogen 
bonding. For  a large-angle rotation of   ${\bi \xi}_k$, 
the  energy needed is 
of order  $a_{\rm HH} \av{|{\bi F}_k^{\rm r}|}_{\rm b}
\cong 29\kBT$. Thus, 
  ${\bi \xi}_k$ and  ${\bi F}_k^{\rm r}$ should be nearly 
parallel in equilibrium, as in the case of  ${\bi \mu}_k$ and 
 ${\bi E}_k^{\rm loc}$. In Fig.10(b), for 
the angle  $ \chi_k= \cos^{-1}(  
{\bi \xi}_k\cdot{\bi F}_k^{\rm r}/ a_{\rm HH}|{\bi F}_k^{\rm r}|)$ 
between these vectors,  its distribution is   again of the 
exponential form,
\bea 
P_\xi^{\rm or}(\cos\chi) &=& 
\av{\delta(\cos\chi- \cos\chi_k)}_{\rm b} \nonumber\\
&\cong &  \exp( B_{\xi} \cos\chi)/Z(B_{\xi} ), 
\ena 
where  $B_\xi=16.6$, so 
  $\av{\cos\chi_k}_{\rm b}=0.93$ 
and $\av{(\cos\chi_k)^2}_{\rm b}=0.87$. The  ${\bi\xi}_k$ 
surely tends to be parallel to  ${\bi F}_k^{\rm r}$.

\subsection{Orientation  time-correlation functions} 

We examine the single-molecule time-correlation functions 
averaged over  the molecules in the bulk. First, we consider 
those  for the local field direction 
${\bi e}_k(t)$ and the dipole direction 
${\bi n}_k(t)$: 
\bea 
C_{\rm loc}(t)&=& \av{{\bi e}_k (t)\cdot{\bi e}_k(0)}_{\rm b},\\
C_{\rm d}(t)&=& \av{{\bi n}_k (t)\cdot{\bi n}_k(0)}_{\rm b}.
\ena 
In Fig.11, we plot  $C_{\rm loc}(t)$ in (a)  and  $C_{\rm d}(t)$ in (b), 
which are  nicely  fitted to the exponential 
form $\propto \exp(-t/\tau_{\rm h})$ with $\tau_{\rm h}=5.0~$ps 
for  $t > 1~$ps. This indicates that   
${\bi n}_k(t)$  moves  with ${\bi e}_k(t)$, 
where ${\bi e}_k(t)$ should be  governed by the hydrogen bond dynamics. 
 In the literature\cite{Rah,Kusa1,Ohmine,Neu,Yip}, 
$C_{\rm d}(t)$  has been found  to decay 
 exponentially 
 for $t>1$ ps at $T\sim 300$ K, 
while it is known to  exhibit  a stretched-exponential 
relaxation in supercooled states\cite{Chen,Stanley}.

\begin{figure}
\includegraphics[width=1\linewidth]{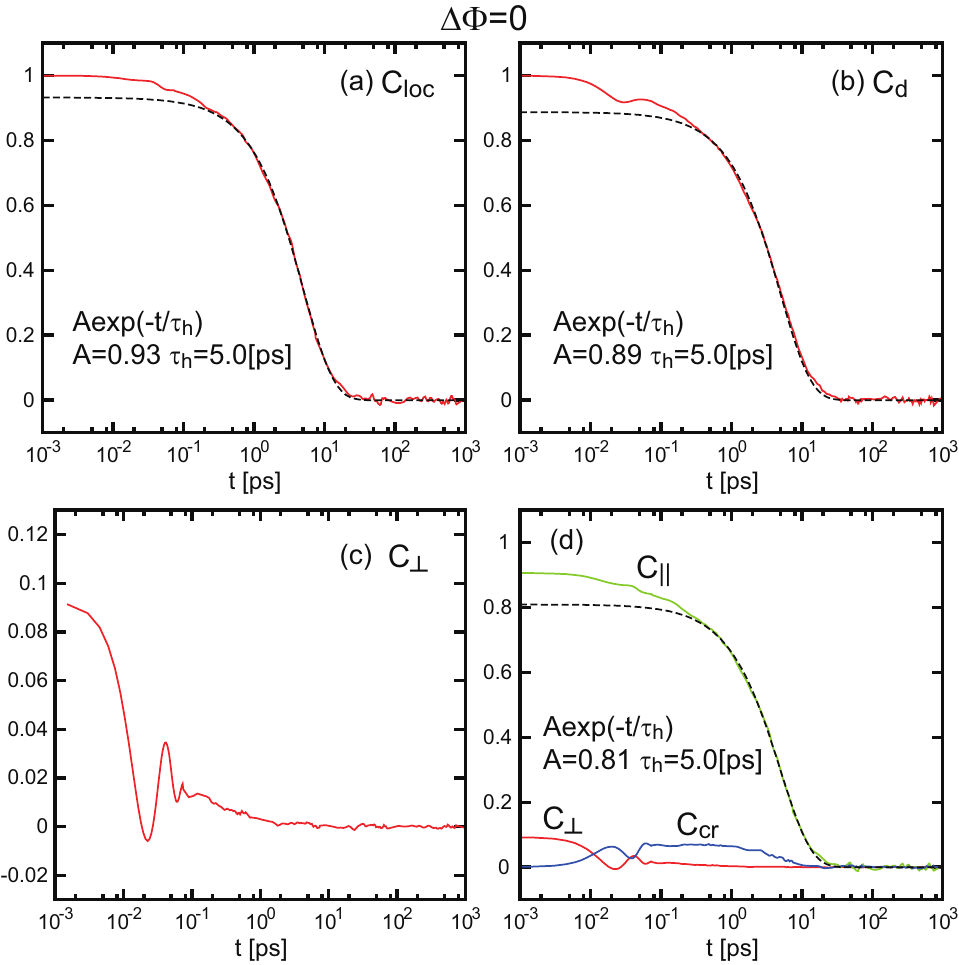}
\caption{  Single-molecule  orientation
time-correlation functions. 
 (a) $C_{\rm loc}(t)$ for the local field direction ${\bi e}_k(t)$ and 
(b) $C_{\rm d}(t)$  for the dipole direction ${\bi n}_k(t)$ (red lines), 
where exponential functions $A \exp(-t/\tau_{\rm h})$ 
are written with common $\tau_{\rm h}=5.0$ ps  (dashed lines). 
 (c) $C_{\perp}(t)$ for the perpendicular 
component ${\bi n}_{k\perp} (t)$ in Eq.(75), which 
exhibits an  oscillatory decay on a timescale of 0.01 ps. 
(d) Decomposition $C_{\rm d}(t)= C_\parallel(t)+ C_\perp(t)  + C_{\rm cr}(t)$ 
using  Eqs.(56)-(58) with exponential fitting for $C_\parallel(t)$ (dashed line). Short-time minimum of $C_{\rm d}(t)$ in (b)
 arises from that of $C_\perp(t)$ in (c).    
} 
\end{figure}

Furthermore, 
we notice that $C_{\rm d}(t)$ has a small local minimum 
at $t=0.03$ ps. To understand this short-time behavior, 
we consider the perpendicular component of ${\bi n}_k(t)$ 
with respect to ${\bi e}_k(t)$ defined by  
\be 
{\bi n}_{k\perp}(t)= {\bi n}_k(t)-[{\bi n}_k(t)\cdot {\bi e}_k(t)]{\bi e}_k(t),\en 
which satisfies ${\bi n}_{k\perp}(t)\cdot {\bi e}_k(t)=0$. 
In Fig.11(c), we plot its single-molecule 
 time-correlation function,
\be 
C_{\perp}(t)= \av{{\bi n}_{k\perp} (t)\cdot{\bi n}_{k\perp}(0)}_{\rm b},  
\en
where $C_\perp(0)= \av{|{\bi n}_{k\perp}|^2}_{\rm b} \cong 0.093$. 
We recognize that ${\bi n}_{k\perp} (t)$ undergoes 
librational motions and $C_\perp(t)$ decays on  a timescale of 0.01 ps  
with a minimum at $t=0.023$ ps. 
In liquid water, similar short-time  behaviors with local minimum and maximum 
have been observed in various time-dependent 
quantities in simulations 
and experiments 
 as manifestation of librational motions\cite{Saito,Hynes,Gei,Laage}. 

To examine the long-time and short-time 
behaviors of $C_{\rm d}(t)$ separately, we also define the parallel 
component $
{\bi n}_{k\parallel}(t)=[{\bi n}_k(t)\cdot {\bi e}_k(t)]{\bi e}_k(t).
$ Then, we  have the decomposition 
$C_{\rm d}(t)= C_\parallel(t)+ C_\perp(t)  + C_{\rm cr}(t)$ with 
\bea 
&&
C_{\parallel}(t)= \av{{\bi n}_{k\parallel} (t)\cdot
{\bi n}_{k\parallel}(0)}_{\rm b},\\
&&\hspace{-1cm} 
C_{\rm cr}(t)= \av{{\bi n}_{k\parallel} (t)\cdot
{\bi n}_{k\perp}(0)}_{\rm b}+  \av{{\bi n}_{k\perp} (t)\cdot
{\bi n}_{k\parallel}(0)}_{\rm b},
\ena  
where $C_\parallel(0)=0.907$ and $C_{\rm cr}(0)=0$. 
In Fig.11(d), we plot these three time-correlation functions. The 
dominant term  $C_\parallel(t)$ has 
 no short-time minimum and 
decays exponentially for $t>1~$ ps. 
Thus, the short-time minimum 
of $C_{\rm d}(t)$ arises from the rapid motions  of 
  ${\bi n}_{k\perp} (t)$. 
The cross contribution 
$C_{\rm cr}(t)$ is at most  0.08   
decaying  exponentially for $t>1~$ps.  


The  scenario in Fig.11  holds also 
for the relative vector  ${\bi \xi}_k(t)$ 
between the two protons in Eq.(69) and its conjugate 
force ${\bi F}_k^{\rm r}(t)$ in Eq.(70).
In  Fig.12(a), displayed are 
\bea 
&&C_{\xi}(t)= 
\av{{{\bi\xi}}_k (t)\cdot{{\bi \xi}}_k(0)}/a_{\rm HH}^2, \\
&&C_{f\xi}(t)= \av{{\bi f}_k^{\rm r} (t)\cdot
{{\bi f}}_k^{\rm r}(0)}, 
\ena 
where ${\bi f}_k^{\rm r} (t)= |{\bi F}_k^{\rm r} (t)|^{-1} 
{\bi F}_k^{\rm r} (t)$ represents the force direction. 
These functions decay  exponentially as 
$ \exp(-t/\tau_\xi)$ with 
 $\tau_\xi=5.8$ ps, which is slightly longer than $\tau_{\rm h}=5.0$ ps  
in Fig.11. To detect librational motions, 
we present the time-correlation function,  
\be 
C_{\xi\perp}(t)=
\av{{{\bi\xi}}_{k\perp} (t)\cdot{{\bi \xi}}_{k\perp}(0)}/a_{\rm HH}^2, 
\en  
 for the vector ${{\bi\xi}}_{k\perp} (t)={\bi \xi}_{k}(t) - 
[{\bi \xi}_{k}(t)\cdot {\bi f}_k^{\rm r}(t)] {\bi f}_k^{\rm r}(t)$ 
 perpendicular to ${\bi f}_k^{\rm r}(t)$. In Fig.12(b), $C_{\xi\perp}(t)$ 
 closely resembles     $C_\perp(t)$ in Fig.11(c), 
  decaying   on a timescale of 0.01 ps.

\begin{figure}
\includegraphics[width=1\linewidth]{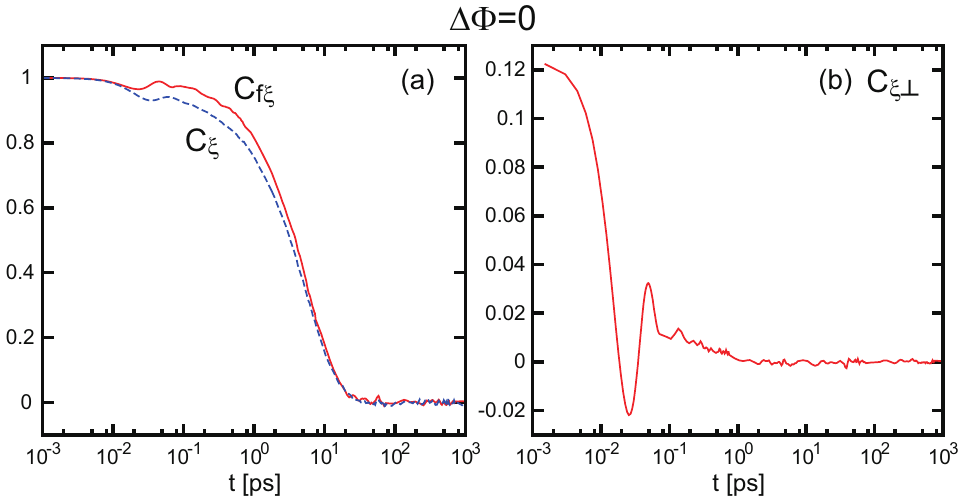}
\caption{Single-molecule  orientation 
time-correlation functions related to 
relative vector ${\bi \xi}_k(t)$ 
between two protons. (a) $C_{f\xi}(t)= \av{{\bi f}_k^{\rm r} (t)\cdot
{{\bi f}}_k^{\rm r}(0)}$  and $C_{\xi}(t)= 
\av{{{\bi\xi}}_k (t)\cdot{{\bi \xi}}_k(0)}/a_{\rm HH}^2$, 
where ${\bi f}_k^{\rm r} (t)= |{\bi F}_k^{\rm r} (t)|^{-1} 
{\bi F}_k^{\rm r} (t)$. These exhibit 
 long-time exponential decay $\propto \exp(-t/\tau_\xi)$ with 
common  $\tau_\xi=5.8$ ps. 
(b) $C_{\xi\perp}(t)$ for 
vector $a_{\rm HH}^{-1}[{\bi \xi}_{k}(t) - 
({\bi \xi}_{k}(t)\cdot {\bi f}_k^{\rm r}(t)) {\bi f}_k^{\rm r}(t)]$ 
 perpendicular to ${\bi f}_k^{\rm r}(t)$, 
 decaying rapidly as  $C_\perp(t)$ in Fig.11(c). 
} 
\end{figure}

\begin{figure}
\includegraphics[width=1\linewidth]{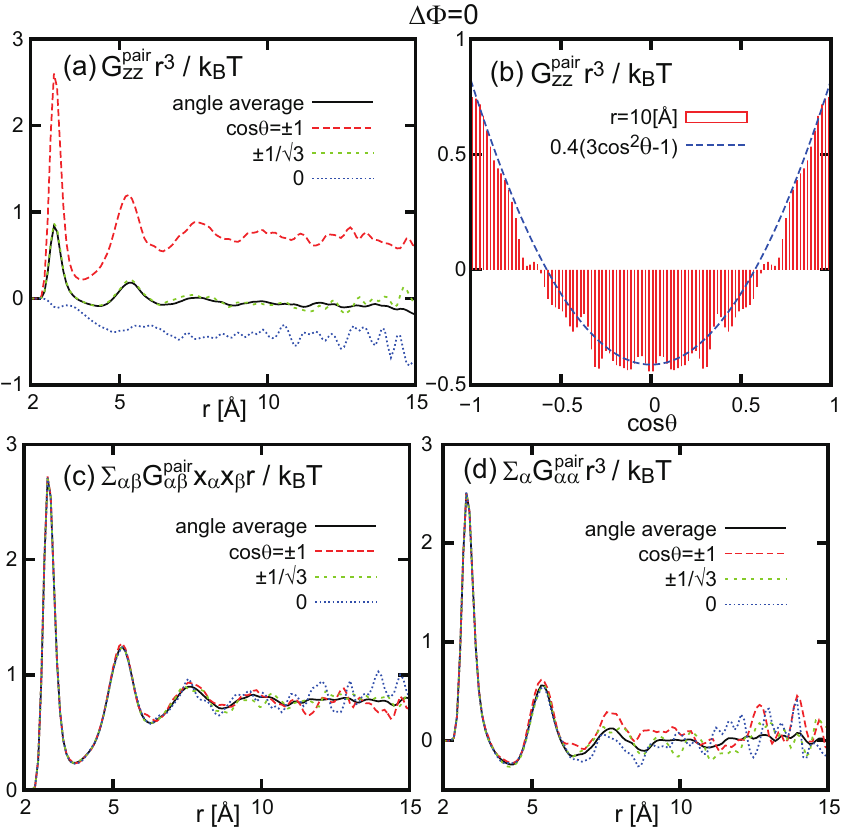}
\caption{ Detection of long-range dipolar correlations in 
$G_{\alpha\beta}^{\rm pair}({\bi r}_0, {\bi r}_0+{\bi r})r^3$ 
for $\Delta\Phi=0$, where the self part is excluded. 
(a) $G_{zz}^{\rm pair}r^3/\kBT$, where 
 $\cos\theta= z/r$ is fixed at $\pm 1, \pm 1/\sqrt{3},$ and 0. 
(b) Its value  at $r=10~{\rm \AA}$, which  
is fitted to  $0.4(3\cos^2\theta-1)$ (broken line). 
(c) Combination $\sum_{\alpha\beta}
 G_{\alpha\beta}^{\rm pair} x_\alpha 
x_\beta r/\kBT$, whose limiting value is about  $ 0.76$. 
(d) Trace $\sum_\alpha G_{\alpha\alpha}^{\rm pair}
r^3/\kBT$ containing  no long-range dipolar contribution.}
\end{figure}

The short-time librational motions are appreciable for the protons 
but are small for the oxygen atoms,  because of the large 
 mass ratio. 
In fact, the mean square displacement of the protons 
$\av{|{\bi r}_{k{\rm H1}}(t+ \Delta t)- {\bi r}_{k{\rm H1}}(t)|^2}_{\rm b}$ 
exhibits a plateau-like behavior 
at  $0.2 $~$\rm \AA^2$ around 
 $\Delta t \sim 0.04 $~ps. This aspect will further be studied in future.

\subsection{Polarization correlations }

\begin{figure}
\includegraphics[width=0.8\linewidth]{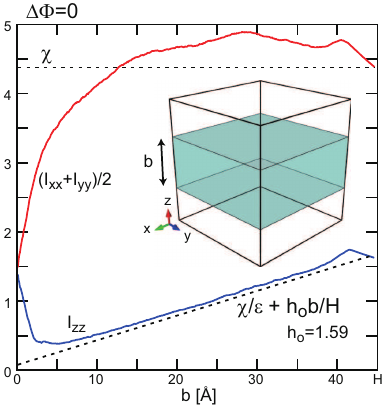}
\caption{$I_{zz}$ and $(I_{xx}+I_{yy})/2$ 
in Eq.(82), which are integrals of 
$G_{\alpha\alpha}({\bi r}_1, {\bi r}_2)/\kBT L^2b$ with respect to ${\bi r}_1$ 
and ${\bi r}_2$  in the plate region with thickness $b$ (light blue region). 
Their behaviors for $b\gg \sigma$ are predicted in 
 Eqs.(83) and (84). For $b>\sigma$, Eq.(83) holds nicely, but 
 Eq.(84) holds only approximately. 
 }
\end{figure} 

We   calculate 
the polarization correlation function  
$G_{\alpha\beta}$ 
in Eq.(30).  There have been some attempts\cite{Kolafa,Tavan,An,Serra,Galli}
 to detect   the long-range dipolar  correlation 
in liquid water, but they have not been based 
on Felderhof's expression (31). 
We aim to detect it unambiguously in agreement with his theory. 
Note that it vanishes 
in the radius-dependent correlation factor $G_{\rm K}(R)$ 
in   Eq.(D6), which has been calculated by many authors. 
We also detect the homogeneous term 
in Eq.(33).

To detect the dipolar correlation, 
we consider   the   pair part 
omitting the self part in  $G_{\alpha\beta}({\bi r}_1,{\bi r}_1+{\bi r})$
(see  Appendix D). For simplicity, 
we use a point-dipole approximation\cite{pair}, 
which is not accurate for  small $r$ but is  
allowable for large $r$. 
In Fig.13,  we  calculate  the product 
 $G_{\alpha\beta}^{\rm pair} r^3$.
From Eqs.(31) or (33),  it should   behave   as 
$(\chi^2/\ve)(3x_\alpha x_\beta/r^2-1)$ for $r\gg \sigma$ with  
 $\chi^2/\ve\cong 0.37$ (from our data of $g_{\rm K}$).
In (a), we plot 
$G_{zz}^{\rm pair}r^3/\kBT$ 
for $\Delta\Phi=0$, where 
 $\cos\theta= z/r=\pm 1, \pm 1/\sqrt{3},$ and 0. 
In (b), the limiting value  can indeed   
be fitted by  $0.4(3\cos^2\theta-1)$. In (c),  the combination  
$\sum_{\alpha\beta}
 G_{\alpha\beta}^{\rm pair}x_\alpha 
x_\beta r/\kBT$ tends to a constant 
about 0.76, while 
its theoretical value is    $2\chi^2/\ve\cong 0.74$. 
In (d), the trace 
$\sum_\alpha G_{\alpha\alpha}^{\rm pair} 
r^3/\kBT$ is displayed, 
which exhibits  no long-range dipolar contribution. 
If it is divided by $n\mu_0^2 r/\kBT$ and is integrated 
with respect to $r$ in the range $0<r<R$, 
we obtain $G_{\rm K}(R)-1$ (see  Eq.(D6)).   
On  all the curves in Fig.13, we can see peaks 
corresponding to nearest neighbor molecules.

To   detect  the homogeneous term in $G_{zz}$ in Eq.(33), 
we integrate  $G_{zz}$ 
over a plate  region 
parallel to the $xy$ plane with thickness $b$,  
where the $z$ coordinate is  in the range $[(H-b)/2,(H+b)/2]$.  
We integrate  $G_{\alpha\alpha}
({\bi r}_1, {\bi r}_2)$ 
with respect to ${\bi r}_1$ and ${\bi r}_2$ 
in this region. The  integrals  are 
divided by ${\kBT L^2b}$ to give  
\be
I_{\alpha\alpha}(b) = \int_{\rm plate} \hspace{-2mm}
 d{\bi r}_1\int_{\rm plate} \hspace{-2mm} d{\bi r}_2
{G_{\alpha\alpha}({\bi r}_1, {\bi r}_2)}/{\kBT L^2b}. 
\en 
In the pancake limit in Eqs.(36) and (37) in     
the lateral periodic boundary condition  ($N_z=1$), we obtain 
\bea 
&&I_{zz}={\chi}/{\ve} + h_{\rm o} {b}/{H} ,\\
&&I_{xx}=I_{yy}=\chi, 
\ena  
which are valid for $b\gg \sigma$.
The first relation also follows from Eq.(B10) in Appendix B. 
In Fig.14,  the curve of 
$I_{zz}$ vs. $b$  nicely 
satisfies Eq.(83) with $h_{\rm o}=1.59$ 
 for $5~{\rm \AA}<b<40~{\rm \AA}$, 
but it exhibits a small peak 
at $b\cong  41~{\rm \AA}$. 
The curve of $(I_{xx}+I_{yy})/2$ tends 
to $\chi$ at $b=H$  but  
exceeds $\chi$ by 10$\%$ 
for $25{\rm \AA}<b<40{\rm \AA}$, where the latter behavior  
is consistent   with   that  of $M_x^{\rm s}$  and  $M_y^{\rm s}$  
in Fig.3(b).   Both $I_{zz}$ and $(I_{xx}+I_{yy})/2$ 
 decrease  for $40{\rm \AA}<b<H$ due to  the depletion 
layers (see  Sec.IIA).


\section{Summary and remarks}

We have  studied the thermal fluctuations of the local field 
and the dipole moments in liquid water 
in applied electric field with    analytic theory 
and molecular dynamics simulation.
In the latter,  we have used  the TIP4P$/$2005 model\cite{Vega}  
and   the 3D Ewald method  \cite{Hautman,Perram,Klapp,Takae,Takae1} 
in a system of $2400$  molecules 
 between metal walls at $z=0$ and $H=44.7~$$\rm \AA$. 
We have  included  the image charges 
and  fixed  the potential difference $\Delta\Phi=HE_{\rm a}$.  
 In the following, we summarize our main results.

(1) 
We have   derived  linear response expressions 
from Eq.(4) in Sec.IIB. In particular, the effective dielectric 
constant $\ve_{\rm eff}$ of film systems is expressed in terms 
of the polarization variance $\av{M_z^2}_0$ 
at zero applied field  in Eq.(17), which is smaller 
than $\av{M_x^2}_0= \av{M_y^2}_0$ by 
the factor $(1+\ell_{\rm w}/H)^{-1}$. 
Here, ${\bi M}$ is the total polarization. 
The $\ell_{\rm w}$ is a surface electric length($\sim~10$ nm)
\cite{Takae1}. 

(2) In Sec.IID, we have presented a theory on  
the polarization correlation function 
$G_{\alpha\beta}({\bi r}_1,{\bi r}_2)$ in Eq.(30) 
in the presence of metal walls.  Together with the 
dipolar term\cite{Felder}, we have found   a homogeneous correlation term 
stemming from   the surface charge fluctuations. 
In Appendix B, this global coupling has been examined  for 
the quantities averaged in the $xy$ plane. 
It  is important for the fluctuations 
of the total polarization $M_z$. However,  it is 
negligible  in a small specimen region in the bulk,  
so we have reproduced    the  Kirkwood-Fr$\ddot{\rm{o}}$hlich   formula 
\cite{Kirk,Fro}.  In Appendix C, we have presented 
another derivation of this formula, where an increase 
of the electrostatic energy is calculated 
for the polarization fluctuations. In Sec.IIIE, we have 
numerically derived  
 the  long-range dipolar term  
 and the homogeneous term 
 in $G_{\alpha\beta}$.

(3) In Sec.IIIA,   we  have examined the fluctuations 
of the local field in applied field. 
 The distribution  $P_{\rm loc}^{\rm ab}(E)$ 
for its  amplitude has been found to be  of  a  Gaussian form  in Eq.(50) 
with a large mean  $E_0 \cong 17~$V$/$nm 
and a relatively small variance. 
This form is mainly due to  the 
hydrogen-bonded molecules, 
as illustrated in Fig.6. 
The joint distribution $P_{\rm loc}(E,\cos\theta)$ 
for  the   amplitude and 
the direction ${\bi e}_k$ of the local field 
is then equal to  the product of $P_{\rm loc}^{\rm ab}(E)$ 
and  the exponential one   
$\propto \exp[B(E)\cos\theta]$, 
where  $B(E)$ depends on $E$ linearly  as in Eq.(49). 
The orientation 
 distribution $P_{\rm loc}^{\rm or}(\cos\theta)$ 
is of the exponential form $\propto \exp(B_{\rm loc}
\cos\theta)$ with $B_{\rm loc}=B(E_{0})$.

(4) In Sec.IIIB,  the  distribution 
 $P_{\rm d}(\cos\theta)$ for the polarization 
direction ${\bi n}_k= \mu_0^{-1} {\bi \mu}_k$ has been 
shown to be of the exponential form and is 
very close to $P_{\rm loc}^{\rm or}(\cos\theta)$ 
for ${\bi e}_k$  in Figs.7 and 8.  
Here, ${\bi n}_k$ and ${\bi e}_k$ 
are nearly parallel from $\mu_0 E_0 
\sim 30 \kBT$. We have also calculated the  distribution 
for the angle $\varphi= 
\cos^{-1}({\bi e}_k\cdot{\bi n}_k)$ between ${\bi n}_k$ and ${\bi e}_k$, 
which  indeed has a sharp peak at $\varphi=0$ in Fig.9.  
In addition, in Sec.IIIC, we have  shown that 
the relative vector ${\bi \xi}_k={\bi r}_{\rm kH1}-{\bi r}_{\rm kH2}$ 
between the two protons tends to be parallel to its conjugate 
force ${\bi F}_k^{\rm r}$ in Eq.(70).  

(5) In Sec.IIID,  we have calculated the single-molecule 
orientation time-correlation 
functions in Figs.11 and 12. Those for 
${\bi e}_k(t)$ and ${\bi n}_k(t)$ 
decay exponentially as $\exp(-t/\tau_{\rm h})$ with 
$\tau_{\rm h}=5.0~$ps for $t >1$ ps, which 
is governed by  the hydrogen bond dynamics. However,  
the perpendicular part of ${\bi n}_{k}(t) $ with respect to 
$ {\bi e}_k(t))$ undergoes librational motions  on a timescale of 0.01 ps. 
We have also calculated those related to 
the relative vector ${\bi \xi}_k(t)$ 
between the two  protons  
to find similar results.

We make some critical remarks.   i)  
We should examine how  the dipole and the local 
field  move together 
in  rotational big jumps in liquid water\cite{Laage,Takae1}. 
 ii) We should consider the molecular polarizability 
due to molecular stretching in strong local field 
\cite{Ab,Raabe}. iii) Ions strongly influence  the local fields  
of the surrounding water molecules \cite{Ka1}, resulting in 
a hydration shell around each ion, 
so this aspect should further be studied. 
iv) In future work, 
we will show that the dipoles ${\bi \mu}_k(t)$ 
are  more aligned along the local field
${\bi E}_k^{\rm loc}(t)$ in supercooled states  than 
at $T=298~$K.  Thus,  the slowing-down of the orientational 
 dynamics in supercooled water 
stems from that of  the hydrogen bond 
dynamics    \cite{Chen,Stanley}.

\begin{acknowledgments}
This work was supported by KAKENHI 
 (Nos. 25610122  and 25000002).  
The numerical calculations were performed on SR16000
at YITP in Kyoto University.
\end{acknowledgments}

\vspace{2mm}
\noindent{\bf Appendix A: Local electric field  and 
polarization density for water}\\
\setcounter{equation}{0}
\renewcommand{\theequation}{A\arabic{equation}}

We explain how the local field and the polarization density 
are defined. In   the TIP4P$/$2005  model\cite{Vega}, 
each water molecule  $k$ forms  a rigid 
isosceles triangle, where  
 $a_{\rm OH}= |{\bi r}_{k{\rm H1}}-{\bi r}_{k{\rm O}}|=
|{\bi r}_{k{\rm H2}}-{\bi r}_{k{\rm O}}|=0.957~{\rm \AA}$ 
with  the HOH angle being  $\theta_{\rm HOH}=104.5^{\circ}$.   
The   charge position 
 M is given by 
$ {\bi r}_{k{\rm M}}={\bi r}_{k{\rm O}}+ a_{\rm OM}{\bi n}_{k}
$   with $a_{\rm OM}=0.1546{\rm \AA}$. 
The   ${\bi n}_k$ is   the unit vector 
from ${\bi r}_{k{\rm O}}$  to the midpoint of  the protons. 
The dipole moment ${\bi \mu}_k$ is written as  Eq.(1). 

The potential $U_{\rm m}$ in Eq.(4) 
depends on the center of mass 
${\bi r}_{k{\rm G}}= 
\frac{8}{9} {\bi r}_{k{\rm O}} 
+ \frac{1}{18} ({\bi r}_{k{\rm H1}} 
+{\bi r}_{k{\rm H2}} )$, the relative vector between the two protons 
 ${\bi\xi}_k ={\bi r}_{k{\rm H1}}-{\bi r}_{k{\rm H2}}$, and 
the dipole moment ${\bi \mu}_k$. 
For  small shifts of the charge positions ${\bi r}_i$  at fixed $E_{\rm a}$, 
 the  potential change $dU_{\rm m}$ is rewritten  as\cite{Takae1} 
\be 
dU_{\rm m}= 
-\sum_k [{\bi F}_k^{\rm e} \cdot d{\bi r}_{k{\rm G}}
+{\bi F}_k^{\rm r}\cdot d{\bi \xi}_k 
+{\bi E}_k^{\rm loc}\cdot d{\bi \mu}_k] .
\en 
The  conjugate electric force to ${\bi r}_{k{\rm G}}$ 
is given by ${\bi F}_k^{\rm e} = q_{\rm H}( {\bi E}_{k{\rm H1}}+ {\bi E}_{k{\rm H2}}
-2{\bi E}_{k{\rm M}})$ and that to  ${\bi \xi}_k$  is 
given in  Eq.(70). 
Conjugate  to ${\bi \mu}_k$  is 
the local electric field,    
\be 
{\bi E}_k^{\rm loc}=
\frac{1}{2}(1+b_{\rm M})({\bi E}_{k{\rm H1}}+ {\bi E}_{k{\rm H2}}) 
-b_{\rm M} {\bi E}_{k{\rm M}}, 
\en 
where $b_{\rm M}$ is a small coefficient given by 
\be 
b_{\rm M}=8 a_{\rm OM}/9 a_{\rm {M{\bar H}}}-1/9= 
0.208. 
\en 
Here,  $a_{\rm {M{\bar H}}}= a_{\rm OH} \cos(\theta_{\rm HOH}/2)-a_{\rm OM}$ 
is  the distance between 
the point M and the midpoint of the protons, while $1/9$ is the 
protons-to-molecule mass ratio.

The polarization density 
${\bi p}({\bi r})$  in Eq.(19) consists of the contributions from 
the constituting molecules as 
${\bi p}({\bi r}) = \sum_k{\bi p}_{k}({\bi r})$. 
For  the TIP4P$/$2005  model, we have\cite{Takae1}
\bea
&&\hspace{-1cm}
{\bi p}_k({\bi r})=  \frac{1}{2} 
[\hat{\delta}({\bi r} ; {\bi r}_{k{\rm H1}},{\bar{\bi r}}_{k{\rm H}})
-\hat{\delta}({\bi r} ; {\bi r}_{k{\rm H2}},{\bar{\bi r}}_{k{\rm H}})]
 q_{\rm H} {\bi \xi}_k 
\nonumber\\
&&\hspace{1cm}+ \hat{\delta}({\bi r} ; {\bar{\bi r}}_{k{\rm H}},
{{\bi r}}_{k{\rm M}})
{\bi \mu}_k  ,
\ena    
where ${\bar{\bi r}}_{k{\rm H}}=({\bi r}_{k{\rm H1}}+{\bi r}_{k{\rm H2}})/2$ is the midpoint of the proton positions, and
$\hat{\delta} ({\bi r} ; {\bi r}_{1},{\bi r}_2) =
\int_0^1 d\lambda~ \delta ( {\bi r}-\lambda {\bi r}_1  
-(1-\lambda){\bi r}_2)$ is the symmetrized $\delta$ function\cite{Onukibook}. 
We  confirm Eq.(19) and obtain 
the dipole and  quadrupole moments\cite{Netz} as   
$\int d{\bi r} {\bi p}_k={\bi \mu}_k$ and 
$\int d{\bi r} ({\bi r}-{\bi r}_{\rm M}) {\bi p}_k=\frac{1}{4}
q_{\rm H}^{-1} {\bi \mu}_k {\bi \mu}_k$  + $\frac{1}{4}
q_{\rm H} {\bi \xi}_k {\bi \xi}_k$. 

\vspace{2mm}
\noindent{\bf Appendix B: Polarization fluctuations 
in the presence of parallel metal walls  
 }\\
\setcounter{equation}{0}
\renewcommand{\theequation}{B\arabic{equation}}
  
We examine  the polarization   fluctuations 
 in highly polar fluids without ions between 
  metal walls. We use    the continuum electrostatics, so 
the polarization density $\bi p$ and the electric field $\bi E$ are smooth  
variables  in the bulk (outside the Stern layers).  
We assume that the cell length  $H$ much exceeds 
 the Stern layer thickness  $d$  for simplicity. 
Then,   the volume  $L^2 (H-2d)$ of the bulk region is 
close to  the cell volume $V=L^2H$. 
 In terms of $\chi$ in Eq.(24) and 
$C$ in Eq.(28),   the  electrostatic energy 
is  written as  
\be
U_{\rm m} = \int_{\rm b} 
 d{\bi r}\bigg[ \frac{|{\bi p}|^2}{2\chi} +\frac{|{\bi E}|^2}{8\pi }
\bigg ]+\int d{\bi r}_\perp 
\frac{{\sigma_0}^2}{2C}  - L^2 \Delta\Phi {\bar \sigma}_0, 
\en  
where $\int_{\rm b} d{\bi r}$ is the integral in 
the bulk region ($d<z<H-d$) and 
$\int d{\bi r}_\perp=\int dxdy$ is the surface integral 
$(0<x,y<L$). 
The first term is the bulk contribution\cite{Felder}, 
the second term is due to the potential drop in  the Stern layers\cite{Beh}, 
and the third term arises from the fixed potential condition. 
In equilibrium,  $U_{\rm m}$ in Eq.(B1) is minimized for 
${\bi p}= P_{\rm b} {\bi e}_z$ and ${\bi E}= 
E_{\rm b} {\bi e}_z$ in the bulk region with 
$\av{\bar\sigma_0}=\ve { E_{\rm b}}/ 4\pi$ (see Eq.(21)).
 
In this appendix, we superimpose deviations homogeneous 
in the $xy$ plane on the above equilibrium averages. 
That is,   we  pick up the Fourier components 
of the deviations with zero wave vector   $(k_x,k_y) = (0,0)$ 
in the $xy$ plane. 
Then, $\delta p_z(z)=p_z-P_{\rm b}$ depends only on $z$. 
As the electric field, 
we consider the fluctuating Poisson electric field 
along the $z$ axis, whose deviation is given by  
\be 
\delta E_z(z) = 4\pi\delta{\bar\sigma}_0-4\pi \delta p_z(z).
\en 
Here, the deviation of the electric induction $\delta D_z(z)=
\delta E_z(z) +4\pi \delta p_z(z)$ is independent of $z$ (in the absence 
of ions). The relation $ \delta p_z(z)=\chi 
\delta E_z(z)$ needs not hold. 
  The fixed potential condition yields the bulk averages, 
\bea
&&\overline{\delta E}_z =
  \int_{\rm b} dz \delta E_z(z)/H= - \delta{\bar\sigma}_0/CH,\\
&&\overline{\delta  p}_z =  \int_{\rm b}  dz \delta { p}_z(z)/H  
 =   (1+\ell_{\rm w}/\ve H ) 
\delta{\bar\sigma}_0,  
\ena 
where $\overline{\delta  p}_z$ is  
 the average of the deviation in the bulk,  
\be 
\overline{\delta p}_z =  \int_{\rm b} d{\bi r}(p_z({\bi r})-P_{\rm b})/V. 
\en 

We then calculate the change  in $U_{\rm m}$ in Eq.(B1) in the 
bilinear (lowest) order in  these  deviations. 
The  bulk  averages  in Eqs.(B3) and (B4) 
($\propto \delta{\bar\sigma}_0)$ give    
\bea
&&
\hspace{-1cm}\delta U_{\rm m}^{(1)}  = 
\frac{V}{2\chi} (\overline{\delta  p}_{\rm b})^2 + 
\frac{V}{8\pi } (\overline{\delta E}_{\rm b})^2 
+ \frac{L^2}{2C} {({\delta{\bar\sigma}}_0)^2} \nonumber\\
&&\hspace{-1cm} =V(1+{\ell_{\rm w}}/{H})
(1+{\ell_{\rm w}}/{\ve H})
 (\delta{\bar\sigma}_0)^2/2\chi  \nonumber\\
&&\hspace{-1cm} =V(1+{\ell_{\rm w}}/{H})
(1+{\ell_{\rm w}}/{\ve H})^{-1}   (
{\overline{\delta  p}}_z)^2/2\chi.
\ena
In the first line, the bulk volume is equated with $V$. 
In the second and third lines,  
${\ell_{\rm w}}/{\ve H}\sim \sigma/H \ll 1$, so this factor can be omitted.
 Namely,  we can  simply set  
$\overline{\delta  p}_z =\delta{\bar\sigma}_0$ and 
$\overline{\delta  E}_z =0$ in Eq.(B6). Next, the inhomogeneous 
deviations $\delta p_z-\overline{\delta 
 p}_z$ 
and   $\delta E_z-\overline{\delta E}_z=-4\pi(\delta p_z-\overline{\delta 
 p}_z)$  yield  
\be 
\delta U_{\rm m}^{(2)}= L^2\int_{\rm b} dz \frac{\ve}{2\chi} 
(\delta p_z-{\overline{\delta  p}}_z)^2. 
\en 
The total change in $U_{\rm m}$ is the sum 
 $\delta U_{\rm m}=\delta U_{\rm m}^{(1)}+\delta U_{\rm m}^{(2)}$ 
and the deviation  $\delta p_z(z)$ obeys the Gaussian distribution 
$\propto \exp(-\delta U_{\rm m}/\kBT)$. 
We can then  calculate  the  pair correlation function in the bulk 
along the $z$ axis:
\be
K(z_1, z_2)= \int d{\bi r}_{1\perp}  \int d{\bi r}_{2\perp}
G_{zz}({\bi r}_1, {\bi r}_2) /\kBT L^2,  
\en 
where $G_{zz}({\bi r}_1, {\bi r}_2)$ is the $zz$ component of the 3D 
 polarization 
correlation function in Eq.(30) with $z_1$ and $z_2$ in the bulk region.  
From Eqs.(B6) and (B7), the 1D  correlation function 
$K(z_1, z_2)$ 
 satisfies the integral equation, 
\be 
\frac{\ve}{\chi} [ K(z_1,z_2)-{\bar K}]  + \frac{1}{\chi} 
(1+\frac{\ell_{\rm w}}{H}) 
{{\bar K}}
= \delta(z_1-z_2).
\en 
Here, ${\bar K}= \int_{\rm b} dz_1 K(z_1,z_2)/H$ 
is the average over $z_1$  independent of $z_2$ (see Eq.(B12)). 
We solve Eq.(B9)  as 
\be 
{K(z_1,z_2)} = \frac{\chi}{\ve}\delta(z_1-z_2) 
+\frac{\chi}{H+\ell_{\rm w}}- \frac{\chi}{\ve H}.
\en 
From Eq.(B12) below, the above 
relation is consistent with Eqs.(33) and (36) with $N_z=1$.
In fact,  the sum of the  last two terms in Eq.(B10) is $h_{\rm o}/H$ 
with $h_{\rm o}$ being given by  Eq.(39). 
 We  have also obtained the coefficient $\chi/\ve$ 
in front of $\delta (z_1-z_2)$. This is  because  we have treated 
the dipolar interaction  as  a short-range one 
(along the $z$ axis) in setting up Eq.(B1).

In the Stern layers,  the microscopic polarization 
 varies  noticeably on the   scale of $d$ (see Fig.2). 
The integral  of its $z$ component in the 
 layers is related to $C$  by   \cite{Takae1} 
\be 
\int_{\rm Stern}\hspace{-4mm}
 d{\bi r}~  p_z=  -\frac{L^2 }{4\pi C}{\bar \sigma}_0
= -\frac{V\ell_{\rm w}}{\ve H} {\bar \sigma}_0,
\en 
which  is equal to $M_z- \int_{\rm b}
 d{\bi r}  p_z$  and is 
consistent with Eq.(B4). Since   $\ell_{\rm w}/\ve  H\ll 1$, 
the  integral of $\delta p_z$ in the Stern layers is 
 much smaller than that in the bulk region. 
Thus, we have   $\int_{\rm b} d{\bi r} \delta p_z
\cong  \delta M_z$, leading to  
\be 
\int_{\rm b} dz_1 K(z_1,z_2)= \frac{\av{(\delta M_z)^2}}{\kBT V}= 
\frac{\chi}{1+\ell_{\rm w}/H},
\en   
which confirms the consistency 
 of Eqs.(23), (39), and (B10).

\vspace{2mm}
\noindent{\bf Appendix C: Dielectric formula for a polar fluid in a sphere   
 }\\
\setcounter{equation}{0}
\renewcommand{\theequation}{C\arabic{equation}}

In the continuum electrostatics,  
 we consider the polarization 
fluctuations in a polar fluid in a sphere  
with radius $R $ much longer than $\sigma$, 
which is  
embedded in an infinite dielectric medium\cite{Fro,Hansen}. 
For the sake of generality, we assume that the dielectric constant of the 
sphere exterior $\ve'$ can be different 
from that in the interior $\ve$. 
If the electric field tends to $E_{\rm b}$ 
along the $z$ axis far from the sphere, 
the  average electric potential $\phi$ is  given by  
\be 
\frac{\phi}{E_{\rm b}z}=-1  -\frac{\ve'-\ve}{2\ve'+\ve} \bigg[\theta(R-r) 
+\theta(r-R)\frac{R^3}{r^3}\bigg] .
\en  
The average 
polarization is given by  ${\bi p}= -{\chi}_{\rm e}(r)\nabla\phi$ in terms of 
the $r$-dependent susceptibility, 
\be 
{\chi}_{\rm e}(r)= \chi \theta(R-r)+\chi'\theta (r-R),
\en 
where $\chi'= (\ve'-1)/4\pi$. 

On the above equilibrium profiles, 
we superimpose  small fluctuations of the 
polarization  and  the electric 
potential  as in Appendix B. They are expressed as 
\bea 
&& \hspace{-5mm} 
\delta{\bi p} = p_{\rm in}  \theta(R-r)  {\bi e}_z
+ p_{\rm ex}\theta(r-R) R^3 \nabla \frac{z}{r^3}, \\
&& \hspace{-5mm}  \delta{\phi}=  -E_{\rm in} z[ \theta(R-r)
+ \theta(r-R) R^3 \frac{1}{r^3}] ,
\ena  
where $p_{\rm in}$ and $ p_{\rm ex}$ represent 
the fluctuation  strength of the polarization 
  in the sphere interior and exterior, respectively.
 Note that 
$\delta\phi$ is continuous at $r=R$ in Eq.(C4). 
We further  require the continuity 
of the electric induction $\delta{\bi D}=\delta{\bi E}+
4\pi \delta{\bi p}$ 
along the surface normal ($\propto {\bi r})$ at $r=R$, 
where $\delta{\bi E}= -{\nabla\delta\phi }$. Then, the strength of 
the electric field 
deviation $E_{\rm in}$  is determined  as 
\be 
E_{\rm in}= -{4\pi} (p_{\rm in}+2 p_{\rm ex})/3.
\en

We use  the electrostatic energy $U_{\rm m}$ in  Eq.(B1) 
 neglecting  the surface capacitance term. 
In the bilinear order of the deviations,  its  incremental change  
 is written as 
 \be 
\delta U_{\rm m} = \int 
 d{\bi r}\bigg[\frac{1}{{2{\chi}_{\rm e} (r)}} {|\delta {\bi p}|^2} +
\frac{1}{8\pi}
{|\delta {\bi E}|^2}\bigg ].  
\en  
From $\nabla\cdot\delta{\bi D}=0$  we have $\int d{\bi r}
\delta{\bi E}\cdot\delta{\bi D}=0$ and 
\be
{\delta U_{\rm m}} = 
\int d{\bi r}~ \frac{1}{2\chi_{\rm e}(r)} 
 \delta{\bi p}\cdot[ 
\delta{\bi p}-\chi_{\rm e} (r) {\delta{\bi E}}].
\en 
Substitution of  Eqs.(C3) and (C4) gives 
\be 
\frac{\delta U_{\rm m}}{v}= 
 \frac{2\ve'+\ve}{2\ve'+1}\cdot \frac{p_{\rm in}^2}{2\chi} +
\frac{2\ve'+1}{3\chi'}
\bigg[p_{\rm ex}+ \frac{\ve'-1}{2\ve'+1} p_{\rm in}\bigg]^2 ,
\en 
where $v=4\pi R^3/3$.  We recognize that 
 $p_{\rm in}$ and the combination 
 $q \equiv p_{\rm ex}+  p_{\rm in}({\ve'-1})/({2\ve'+1})  $ 
are independent  Gaussian random variables obeying   the 
distribution $\propto \exp( -\delta U_{\rm m}/\kBT)$. 
In the linear response regime, 
 we can set 
$\delta M_{z}^{\rm s} = \int_{r<R} d{\bi r}\delta p_z= 
vp_{\rm in}$ to obtain  
\be 
 \av{(\delta M_z^{\rm s})^2}= v\kBT \chi (2\ve'+1)/(2\ve'+\ve), 
\en 
which is well-known in 
 the  literature \cite{Leeuw,Hansen,Sm}.  If $\ve'=\ve$, 
we reproduce Eq.(40) for $v\ll V$. 

Using simple arguments,  Fr$\ddot{\rm{o}}$hlich \cite{Fro} 
obtained the first term in Eq.(C8) for $\ve'=\ve$. 
To reproduce his result, let us set   $q=0$ at fixed $p_{\rm in}$. 
Then, $E_{\rm in}= -{4\pi} p_{\rm in}/(2\ve'+1)=  p_{\rm ex}/\chi'$ 
from Eq.(C5) so that   
 $\delta{\bi p}- {\chi}_{\rm e} \delta{\bi E}$ 
vanishes for $r>R$ 
and is equal to $[(\ve+2\ve')/(2\ve'+1)]p_{\rm in}{\bi e}_z$ for $r<R$. 
Thus, the outer region is in a local equilibrium state for $q=0$, 
which was assumed in the reaction field theory\cite{Onsager,Neu}. 
Now,  from Eq.(C7),  the first term in Eq.(C8)  readily follows.

\begin{figure}[t]
\includegraphics[width=0.96\linewidth]{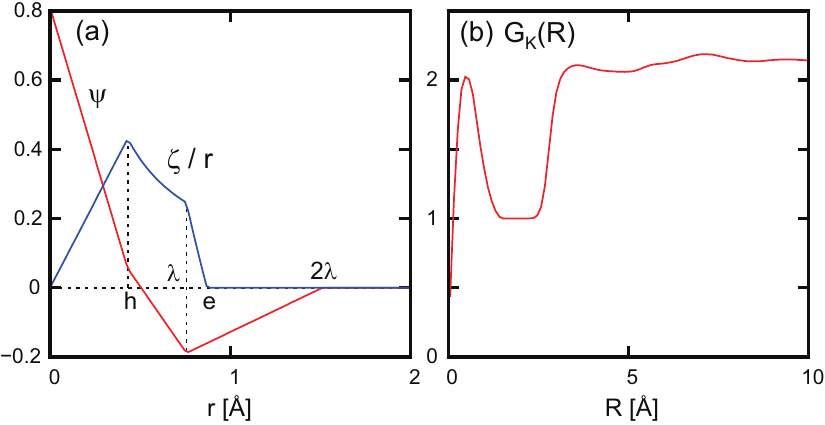}
\caption{   (a) Functions $\psi(r)$ 
and $\zeta(r)$ in Eqs.(D4) and (D5) 
vs. $r$, where $2\lambda=a_{\rm HH}$ (HH distance) and 
$h= a_{\rm M{\bar H}}$ (distance between M and 
the proton midpoint). 
(b)  Radius-dependent correlation factor 
 $G_{\rm K}(R)$ in Eq.(D6) vs. $R$, 
which tends to the correlation factor $g_{\rm K}$ for 
$R >2~{\rm \AA}$. It has a maximum at $r=2\lambda/3$ 
due to the unique shape of the water molecules. 
 }
\end{figure}

\vspace{2mm}
\noindent{\bf Appendix D: Polarization correlation function  
 }\\
\setcounter{equation}{0}
\renewcommand{\theequation}{D\arabic{equation}}

Here, we calculate the self part of the  
polarization correlation function in Eq.(30).
In terms of  ${\bi p}_{k}({\bi r})$ in Eq.(A4), it is expressed as    
\bea 
&&\hspace{-2cm} 
G_{\alpha\beta}({\bi r}_1,{\bi r}_2) =  \sum_{k,\ell} 
\av{p_{k\alpha}({\bi r}_1)p_{\ell\beta}({\bi r}_2)}\nonumber\\
&&\hspace{-0.5cm}= G_{\alpha\beta}^{\rm self}({\bi r}_1,{\bi r}_2)
+G_{\alpha\beta}^{\rm pair}({\bi r}_1,{\bi r}_2),
\ena 
where  we assume  $E_{\rm a}=0$.  
 In the second line, 
we divide $G_{\alpha\beta}$ into the  self part 
$G_{\alpha\beta}^{\rm self}$ with $k=\ell$  
and the pair part $G_{\alpha\beta}^{\rm pair}$   
  with $k\neq \ell$.

For point-dipole models, we 
would have $G_{\alpha\beta}^{\rm self}
({\bi r}_1,{\bi r}_2)= n\mu_0^2\delta_{\alpha\beta}
\delta({\bi r}_1-{\bi r}_2)/3$. 
For the TIP4P$/$2005  model\cite{Vega}, the self part 
 depends on the following  molecular lengths,
\be
h=a_{\rm {M{\bar H}}}
=0.431~{\rm \AA}, 
\quad 
\lambda =a_{\rm {H{ H}}}/2 
= 0.757~{\rm \AA},
\en 
where $h$  is the distance between 
the point M  and the midpoint of the two protons,
and $\lambda $ is half of the distance between the two protons. 
  The   averages over the orientations 
of ${\bi n}_k$ and ${\bi \xi}_k$   give 
\be 
G_{\alpha\beta}^{\rm self}({\bi r}_1,{\bi r}_2)
= \frac{n\mu_0^2}{2\pi h^2}\bigg[\bigg(\psi+\frac{3\zeta}{4r}\bigg)
 \frac{{{x}}_\alpha{ x}_\beta}{r^4}  - \zeta\frac{\delta_{\alpha\beta}}{4r^3}
 \bigg] ,
\en 
where ${\bi r}= {\bi r}_2-{\bi r}_1$. 
We define 
$\psi=\psi(r)$ and $\zeta=\zeta(r)$ as   
\bea 
&&\hspace{-1cm}
\psi(r)= (h-r)\theta(h-r) + (\lambda-r)\theta(\lambda-r) \nonumber\\
&&- ({2\lambda-r})\theta(2\lambda-r)/4,\\
&&\hspace{-1cm}
\zeta(r)=
-(h^2-r^2)\theta(h-r)- (\lambda^2-r^2)\theta(\lambda-r)\nonumber\\
&&+(e^2-r^2) \theta(e-r), 
\ena 
where $\theta(u)$ is the step function 
being equal to 1 for $u>0$ and 
0 for $u\le 0$ and $e=\sqrt{h^2+\lambda^2}=0.871~{\rm \AA}$. 
The self part is nonvanishing only  for $r<2\lambda=1.51~{\rm \AA}$, 
where the upper bound is the H-H distance  $a_{\rm{HH}}$. 
As shown in Fig.15(a),  $\psi(r)>0$ for 
$r<2\lambda/3$ and $\psi(r)<0$ for 
$2\lambda/3<r<2\lambda$ leading to $\int_0^{2\lambda} dr \psi(r)= h^2/2$, 
while $\zeta (r)\ge 0$ for any $r$. 
Therefore, the integral  of $ G_{\alpha\beta}^{\rm self}({\bi r}_1,{\bi r}_1+
{\bi r})$ with respect to $\bi r$  
is equal to $ n\mu_0^2\delta_{\alpha\beta}/3$, as ought to be the case. 

Following the previous authors \cite{Rah,Neu,Kusa,Yip,Beren}, 
we calculate  the radius-dependent correlation factor defined by  
\be 
G_{\rm K}(R)\equiv \int_{r<R} d{\bi r}\sum_\alpha 
G_{\alpha\alpha}({\bi r}_1,{\bi r}_1+{\bi r})/n\mu_0^2. 
\en 
In Fig.15(b), we plot $G_{\rm K}(R)$ vs the radius  $R$ using 
 the self part in Eq.(D3)  and 
a point-dipole approximation 
for the pair part\cite{pair}. 
 We confirm the plateau behavior  $G_{\rm K}(R)\cong  g_{\rm K}= 2.14$ 
for  $\sigma\ll R\ll V^{1/3}$.


\begin{thebibliography}{0}
\bibitem{Fro}
H. Fr$\ddot{\rm{o}}$hlich, {\it Theory of dielectrics} (Oxford University Press, Oxford, 1949).

\bibitem{Onsager}
L. Onsager, J. Am. Chem. Soc. {\bf 58}, 1486 (1936).

\bibitem{Kirk}
J. G. Kirkwood, J. Chem. Phys. {\bf 7}, 911 (1939).

\bibitem{Harris}
F. E. Harris and B. J. Alder, J. Chem. Phys. {\bf 21}, 1031 (1953).

\bibitem{Deu}
G. Nienhuis and  J. M.  Deutch, J. Chem. Phys. {\bf 55}, 4213 (1971).

\bibitem{Felder}
B. U. Felderhof, J. Chem. Phys. {\bf 67}, 493 (1977);
J. Phys. C: Solid State Phys. {\bf 12}, 2423 (1979).

\bibitem{Wer}
M. S. Wertheim, Ann. Rev. Phys. Chem. {\bf 30}, 471 (1979).

\bibitem{Stell}
G. Stell, G. N. Patey, and J. S.  H$\rm{\o}$ye, Adv. Chem. Phys. {\bf 48}, 183 (1981).
  
\bibitem{Sm}
J. W. Perram  and E. R. Smith, J. Stat. Phys. {\bf  46}, 179 (1987).
\bibitem{Hansen}
V. Ballenegger and J.-P. Hansen, Europhys. Lett. {\bf 63}, 381 (2003);
R. Blaak and J.-P. Hansen, J. Chem. Phys.  {\bf  124}, 144714 (2006).

\bibitem{Netz}
D. J.  Bonthuis, S. Gekle, and R. R. Netz, Langmuir {\bf 28}, 7679 (2012).
 



\bibitem{Allen}
M. P. Allen and D. J. Tildesley, {\it Computer Simulation of Liquids} (Clarendon Press, Oxford, 1987).
\bibitem{Rah}
A.  Rahman and F. H. Stillinger, J. Chem. Phys. {\bf 55}, 3336 (1971).

\bibitem{Leeuw}
S. W. de Leeuw, J. W. Perram, and E. R. Smith, Proc. R. Soc. Lond. A {\bf 373}, 27 (1980);
Ann. Rev. Phys. Chem. {\bf 37}, 245 (1986).

\bibitem{Neu}
M.  Neumann, Mol. Phys.  {\bf 50}, 841 (1983); J. Chem. Phys. {\bf 85}, 1567 (1986).

\bibitem{Yip}
J. Anderson, J. J. Ullo, and S. Yip, J. Chem. Phys. {\bf 87}, 1726 (1987).

\bibitem{Kusa}
P. G. Kusalik, J. Chem. Phys. {\bf 93}, 3520 (1990).

\bibitem{Beren}
D. van der Spoel, P. J. van Maaren, and H. J. C. Berendsen, J. Chem. Phys. {\bf 108}, 10220 (1998).

\bibitem{St}
P. H$\rm{\ddot o}$chtl, S. Boresch, W. Bitomsky, and O. Steinhauser, J. Chem. Phys. {\bf 109}, 4927 (1998).

\bibitem{Vega}
J. L. F. Abascal and C. Vega, J. Chem. Phys. {\bf 123}, 234505 (2005).


\bibitem{Raabe}
G. Raabe and R. J. Sadus, J. Chem. Phys. {\bf 134}, 234501 (2011).


\bibitem{Ab}
M. A. Gonz${\acute{\rm a}}$lez and J. L. F. Abascal, J. Chem. Phys.  {\bf 135}, 224516 (2011).





\bibitem{Hautman}
J. Hautman, J. W. Halley, and Y.-J. Rhee, J. Chem. Phys. {\bf 91}, 467 (1989).

\bibitem{Perram}
J. W. Perram and M. A. Ratner, J. Chem. Phys. {\bf 104}, 5174 (1996).

\bibitem{Klapp}
S. H. L. Klapp, Mol. Simul. {\bf 32}, 609 (2006).

\bibitem{Takae}
K. Takae and A. Onuki, J. Chem. Phys. {\bf 139}, 124108 (2013).

\bibitem{Takae1}
K. Takae and A. Onuki, J. Phys. Chem. B {\bf 119}, 9377 (2015).

\bibitem{Yeh}
I.-C. Yeh and M. L. Berkowitz, J. Chem. Phys. {\bf 110}, 7935 (1999);
{\it ibid}. {\bf 111}, 3155 (1999).

\bibitem{Hender}
P. S. Crozier, R. L. Rowley, and D. Henderson, J. Chem. Phys.  {\bf 113}, 9202 (2000).

\bibitem{Voth}
M. K. Petersen, R. Kumar, H. S. White, and G. A. Voth, J. Phys. Chem. C {\bf 116}, 4903 (2012).

\bibitem{Madden1}
A. P. Willard, S. K. Reed, P. A. Madden, and D. Chandler, Faraday Discuss. {\bf 141}, 423 (2009).

\bibitem{Sprik}
J. I. Siepmann and M. Sprik, J. Chem. Phys. {\bf 102}, 511 (1995).  



\bibitem{Sayka}
J. D. Smith, C. D. Cappa, K. R. Wilson, R. C. Cohen, P. L. Geissler, and R. J. Saykally, PNAS {\bf 102}, 14171 (2005). 

\bibitem{Dellago}
B. Reischl, J. K$\ddot{\rm o}$finger, and C. Dellago, Mol. Phys. {\bf 107}, 495 (2009).

\bibitem{Ka1} 
B. Sellner, M. Valiev, and S. M. Kathmann, J. Phys. Chem. B  {\bf 117}, 10869 (2013). 
In  analysis in this paper, the local  field is set equal 
to $E_0{\bi \nu}_k$, where ${\bi \nu}_k$ 
is a unit vector determined for each molecule $k$. 
The distribution in Eq.(45) 
is  the orientation average of the Gaussian 
$\av{{\exp[-|{\bi E}- E_0{\bi \nu_{\bi k}}|^2/2s_0]}/(2\pi s_0)^{3/2}}$
over  ${\bi \nu}_k$.  





\bibitem{Ohmine}
H. Tanaka and I. Ohmine, J. Chem. Phys. {\bf 87}, 6128 (1987);
I. Ohmine, J. Phys. Chem. {\bf 99}, 6767 (1995).

\bibitem{Kusa1}
I. M. Svishchev  and P. G. Kusalik, J. Chem. Soc. Faraday Trans. {\bf 90}, 1405 (1994). 

\bibitem{Gei}
J. D. Eaves, A. Tokmakoff, and P. L. Geissler, J. Phys. Chem. A {\bf 109}, 9424 (2005).


\bibitem{Saito} 
T. Yagasaki, J. Ono, and S. Saito, J. Chem. Phys. {\bf 131}, 164511 (2009).

\bibitem{Laage}
D. Laage, G. Stirnemann, F. Sterpone, R. Rey, and J. T. Hynes, Annu. Rev. Phys. Chem. {\bf 62}, 395 (2011).


\bibitem{Hynes}
J. Petersen, K. B. M${\rm {\o}}$ller,  R. Rey, and J. T. Hynes, J. Phys. Chem. B {\bf 117}, 4541 (2013). 




\bibitem{Onukibook} A. Onuki, 
{\it Phase Transition Dynamics} 
(Cambridge University Press, Cambridge, 2002).



\bibitem{Limmer}
D. T. Limmer, C. Merlet, M. Salanne, D. Chandler, P. A. Madden, R. van Roij, and B. Rotenberg,
Phys. Rev. Lett. {\bf 111}, 106102 (2013).




\bibitem{Beh}
S. H. Behrens and M. Borkovec, J. Phys. Chem. B {\bf 103}, 2918 (1999); 
S. H. Behrens and D. G. Grier, J. Chem. Phys. {\bf 115}, 6716 (2001). 


\bibitem{Bazant} 
M. Z. Bazant, M. S. Kilic, B. D. Storey, and A. Ajdari, Adv. Colloid Interface Sci. {\bf 152}, 48 (2009).


\bibitem{Ah}
A. Aharony and M. E. Fisher, Phys. Rev. A {\bf 8}, 3323 (1973). 



\bibitem{surface} 
The  integral of the dipolar part with respect 
to $\bi r$ in the spheroid  can be transformed into the 
surface integral, where we have 
$\psi_\ell(|{\bi r}-{\bi r}'|)\cong |{\bi r}-{\bi r}'|^{-1}$ 
with  ${\bi r}$ being on the surface and  ${\bi r}'$ being 
in the interior far from the surface.  
Then, we obtain  Eqs.(36) and (37). 


\bibitem{Landau}
L. D. Landau and E. M. Lifshitz, {\it Electrodynamics of Continuous Media} (Pergamon, 1984) Chap II.


\bibitem{comment2} 
In the lateral periodic boundary condition 
in  simulation, 
$I({\bi r}) \equiv 
\nabla_\alpha\nabla_\beta\psi_\ell(r)$ in Eq.(33)
is  replaced by 
$J ({\bi r})\equiv \sum_{{\bi m}_\perp}\nabla_\alpha\nabla_\beta\psi_\ell
(|{\bi r}- L {\bi m}_\perp|)$,  
where ${\bi m}_{\perp}=(m_x, m_y,0)$ as in Eq.(9). 
Then, the integral of 
$J ({\bi r}) $ on the wall $0<x,y<L$  becomes 
that of $I({\bi r}) $ in the  plane ${-\infty}<x,y<\infty$.


\bibitem{Sengers} 
D. P. Fernandez, A. R. H. Goodwin, E. W. Lemmon, J. M. H. Levelt Sengers, and R. C. Williams,
J. Phys. Chem. Ref. Data {\bf 26}, 1125 (1997).


\bibitem{Luzar} 
A. Luzar and D. Chandler, Phys. Rev. Lett. {\bf 76}, 928 (1996).


\bibitem{Zie}
J.  Zielkiewicz, J. Chem. Phys. {\bf 123}, 104501 (2005). 


\bibitem{Kumar}
R. Kumar, J. R. Schmidt, and J. L. Skinner, J. Chem. Phys. {\bf 126}, 204107 (2007).

\bibitem{Rao} 
D. Prada-Gracia, R. Shevchuk, and F. Rao,  J. Chem. Phys. {\bf 139}, 084501 (2013). 



\bibitem{comment3} 
 For each reference $k$ in the bulk,  
the surrounding molecules  
in the region $ |{\bi r}_{\ell{\rm G}}-{\bi r}_{k{\rm G}}|<4.4~{\rm \AA}$
amount to  11.5 and $17\%$ of them are neither 
first  nor second nearest ones 
on the average, where   ${\bi r}_{k{\rm G}}$ and ${\bi r}_{\ell{\rm G}}$ 
are  the centers  of mass. 

\bibitem{Koga}
Y. He, G. Sun, K. Koga, and L. Xu, Sci. Rep. {\bf 4}, 6596  (2014).

\bibitem{Chen} 
F. Sciortino,  P. Gallo,  P. Tartaglia,  and S.-H. Chen, Phys. Rev. E, {\bf 54}, 6331 (1996). 
\bibitem{Stanley}
P. Kumar, G. Franzese, S. V. Buldyrev, and H. E. Stanley, Phys. Rev. E {\bf 73}, 041505 (2006).


\bibitem{Kolafa} 
J. Kolafa and I. Nezbeda, Mol. Phys. {\bf 98}, 1505 (2000).

\bibitem{Tavan}
G. Mathias and P. Tavan, J. Chem. Phys. {\bf 120}, 4393 (2004).

\bibitem{An} 
J. M. P. Kanth, S. Vemparala, and R. Anishetty, Phys. Rev. E {\bf 81}, 021201 (2010).

\bibitem{Serra} 
D. C. Elton and M.-V. Fern$\acute{\rm a}$ndez-Serra, J. Chem. Phys. {\bf 140}, 124504 (2014).
 
\bibitem{Galli} 
C. Zhang and G. Galli, J. Chem. Phys. {\bf 141}, 084504 (2014).

\bibitem{pair} 
In Figs.13 and 15, the pair part $G_{\alpha\beta}$ 
is approximated as 
$G_{\alpha\beta}^{\rm pair}({\bi r}_1,{\bi r}_1+{\bi r})
= \sum_{k\neq \ell}\av{{\bi \mu}_k \cdot{\bi \mu}_\ell 
\delta({\bi r}- {\bi r}_{k\ell})}_{\rm b} /V $ 
with ${\bi r}_{k\ell}= \frac{1}{4}( {\bi r}_{k{\rm H1}}+ 
{\bi r}_{k{\rm H2}} -{\bi r}_{\ell{\rm H1}}-{\bi r}_{\ell{\rm H2}}) 
+\frac{1}{2}({\bi r}_{k{\rm O}}- {\bi r}_{\ell{\rm O}})$. 


\end{thebibliography}
\end{document}